\documentclass[a4paper,fleqn,usenatbib]{mnras}

\usepackage{newtxtext,newtxmath}
\usepackage[T1]{fontenc}
\usepackage{ae,aecompl}

\usepackage{amsmath}
\usepackage{amssymb}
\usepackage[section]{placeins}
\usepackage{color}
\usepackage[usenames,dvipsnames]{xcolor}

\usepackage{mathrsfs}
\usepackage{mathtools}
\usepackage{natbib}
\usepackage{subfigure}
\usepackage{psfrag}
\usepackage{layouts}
\usepackage{upgreek}
\usepackage{graphicx}
\usepackage{soul}
\usepackage{caption}
\usepackage{pdflscape}
\usepackage{longtable}
\usepackage{supertabular}
\usepackage{afterpage}
\usepackage{enumitem}

\def\aap{A\&A}                % Astronomy and Astrophysics
\def\aapr{A\&A~Rev.}          % Astronomy and Astrophysics Reviews
\def\aj{AJ}                   % Astronomical Journal
\def\apj{ApJ}                 % Astrophysical Journal
\def\apjl{ApJ}                % Astrophysical Journal, Letters
\def\apjs{ApJS}               % Astrophysical Journal, Supplement
\def\pasp{PASP}
\def\mnras{MNRAS}%
\def\nat{Nature}%
\def\pasj{PASJ}%              % Publications of the Astronomical Society of Japan

\title[The life cycles of Be viscous decretion discs]{The life cycles of Be viscous decretion discs: Time-dependent modelling of infrared continuum observations}

\author[R. G. Vieira, A. C. Carciofi, J. E. Bjorkman, Th. Rivinius, D. Baade and L. R. R\'imulo]{
\parbox{\textwidth}{R. G. Vieira$^{1}$\thanks{E-mail: rg.vieira@gmail.com},
A. C. Carciofi$^{1}$,
J. E. Bjorkman$^{2}$,
Th. Rivinius$^{3}$,
D. Baade$^{4}$
and L. R. R\'imulo$^{1}$} \vspace{0.4cm}\\
\parbox{\textwidth}{
$^{1}$Instituto de Astronomia, Geof\'isica e Ci\^encias Atmosf\'ericas,
      Universidade de S\~ao Paulo, Rua do Mat\~ao 1226, Cidade Universit\'aria, 05508-900 S\~ao Paulo, SP, Brazil\\
$^{2}$Ritter Observatory, Department of Physics \& Astronomy, University of Toledo, Toledo, OH 43606, USA\\
$^{3}$ESO - European Organisation for Astronomical Research in the Southern Hemisphere, Casilla 19001, Santiago 19, Chile\\
$^{4}$ESO - European Organisation for Astronomical Research in the Southern Hemisphere, Karl-Schwarzschild-Str. 2, 85748 Garching bei M\"unchen, Germany
}}

\date{Accepted 2016 October 3. Received 2016 September 26; in original form 2016 May 16}

\pubyear{2016}

\begin{document}
\label{firstpage}
\pagerange{\pageref{firstpage}--\pageref{lastpage}}
\maketitle

\begin{abstract}
We apply the viscous decretion disc (VDD) model to interpret the infrared disc continuum emission of 80 Be stars observed in different epochs. In this way, we determined 169 specific disc structures, namely their density scale, $\rho_0$, and exponent, $n$. We found that the $n$ values range mainly between $1.5$ and $3.5$, and $\rho_0$ varies between $10^{-12}$ and $10^{-10}\,\mathrm{g\,cm^{-3}}$, with a peak close to the lower value. 
Our large sample also allowed us to firmly establish that the discs around early-type stars are denser than in late-type stars. Additionally, we estimated the disc mass decretion rates and found that they range between $10^{-12}$ and $10^{-9}\,\mathrm{M_{\odot}\,yr^{-1}}$. These values are compatible with recent stellar evolution models of fast-rotating stars.
One of the main findings of this work is a correlation between the $\rho_0$ and $n$ values. In order to find out whether these relations can be traced back to the evolution of discs or have some other origin, we used the VDD model to calculate temporal sequences under different assumptions for the time profile of the disc mass injection. The results support the hypothesis that the observed distribution of disc properties is due to a common evolutionary path.
In particular, our results suggest that the timescale for disc growth, during which the disc is being actively fed by mass injection episodes, is shorter than the timescale for disc dissipation, when the disc is no longer fed by the star and dissipates as a result of the viscous diffusion of the disc material.
\end{abstract}


%%%%%%%%%%%%%%%%%%%%%%%%%%%%%%%%%%%%%%%%%%%%%%%%%%

%%%%%%%%%%%%%%%%% BODY OF PAPER %%%%%%%%%%%%%%%%%%

% INTRODUCTION___________________________________

\section{Introduction}
\label{sect:introduction}

% Be stars
Be stars are early-type objects with ionized gaseous discs. The flattened geometry of these circumstellar structures was first suggested by \citet{struve1931}, which was confirmed many years later by interferometric observations \citep[e.g.,][]{dougherty1992, stee1995, quirrenbach1997}. Based on the observed fraction of Be-shell stars, \citet{porter1996} estimated a value of $5^{\circ}$ for the disc opening angle, while \citet{quirrenbach1997} found an upper limit of $20^{\circ}$ from interferometric and spectropolarimetric observations. In particular, \citet{wood1997} found a disc opening angle of $2.5^{\circ}$ using spectropolarimetry. The apparent inconsistency among these determinations can be explained by the disc flaring at larger radii, and the fact that distinct observational techniques probe different disc regions \citep[e.g.,][]{carciofi2011}. Based on the study of Fe~II shell line profiles, \citet{hanuschik1996} showed that the observations are consistent with a rotationally supported geometrically thin disc in vertical hydrostatic equilibrium.

% Waters contribution
\citet{waters1987} determined the disc density structure of 54 Be stars based on their \textit{IRAS} infrared (IR) flux excesses. By assuming an outflowing disc model with a power law density profile, a fixed opening angle of $15^{\circ}$, and a radial velocity at the disc base of $5\,\mathrm{km\, s^{-1}}$, they constrained the density slope exponent to be in the range between $2$ and $3.5$, and the mass loss rates, $\dot{M}$, to typically lie between $10^{-9}$ and $10^{-7}\,\mathrm{M_{\odot}\,yr^{-1}}$. Based on the upper limit of their observed $\dot{M}- L_{\star}$ distribution, these authors also suggested a $\dot{M}$ regime transition at $\sim$$10^3\,\mathrm{L_{\odot}}$. These results still remain as a reference for typical Be disc properties \citep[e.g.,][]{granada2013}. Such $\dot{M}$ values are usually much larger than those found for normal B stars, which range from $10^{-11}$ to $3\times 10^{-9}\,\mathrm{M_{\odot}\,yr^{-1}}$ \citep{snow1981}

% VDDM new paradigm
In the past decade, major steps forward were achieved in our understanding of Be star discs \citep[see the recent review paper by][]{rivinius2013}. The viscous decretion disc (VDD) model \citep{lee1991} became the new paradigm for the interpretation of Be stars observations \citep{carciofi2011}. For a handful of objects, detailed modeling using the VDD model has successfully reproduced multi-technique observations \citep[e.g.,][]{carciofi2006b, tycner2008, jones2008, carciofi2009, klement2015}.
% Haubois 2012
In addition to static models, the VDD model also allows the study of the dynamical evolution of the disc. For example, \citet{carciofi2012} successfully described the light curve of 28~CMa using the time-dependent VDD model, providing the first determination of the viscosity parameter for a Be star disc. In particular, \citet{haubois2012} showed how the observed light curves are affected by the mass injection rate history, and related the density exponent to the disc dynamical state. They found that steep radial density profiles typically correspond to the disc build-up phase, while flatter density slopes are usually related to the disc dissipation. Based on smoothed particle hydrodynamic (SPH) simulations, \citet{okazaki2002} and \citet{panoglou2016} demonstrated that flatter density profiles may also be related to the accumulation effect caused by binary interaction.

% pseudo-photosphere model
To relate the VDD dynamical predictions to observations, it is necessary to compute the radiative transfer in the circumstellar environment. State-of-the-art codes, such as \texttt{HDUST} \citep{carciofi2006, carciofi2008} and \texttt{BEDISK} \citep{sigut2007}, are capable of solving the three-dimensional non-LTE radiative transfer problem, computing the disc continuum emission, polarization and line profiles for the VDD model. In particular, \citet{vieira2015} developed the pseudo-photosphere model, which consists of a simple and accurate semi-analytic formulation to compute the disc continuum emission. It separates the disc into two components: an inner optically thick region (the pseudo-photosphere) and an outer optically thin region. Such simplification allows one to derive analytical expressions for the disc flux and spectral slope, which were calibrated and validated by the full radiative transfer calculations of \texttt{HDUST}. Based on this new approach, \citet{vieira2015} showed that the spectral slope is mainly determined by the density radial slope and disc flaring exponent, and is rather insensitive to the base density and disc inclination. As a first application, \citet{vieira2015} fitted the \textit{IRAS} flux excess of a sub-sample of stars presented by \citet{waters1987} to constrain the disc parameters. Among other results, they found that the mass decretion rate derived for this small sample lies between $10^{-12}$ and $10^{-9}\,\mathrm{M_{\odot}\,yr^{-1}}$, which is at least two orders of magnitude smaller than the values previously found by \citet{waters1987}. This difference arises from the application of the VDD model rather than \citeauthor{waters1987}'s outflowing disc model. For a viscosity parameter $\alpha=1$, the VDD model predicts a radial velocity at the disc base of $v_{\varpi} \sim 10^{-3}\, c_{\mathrm{s}}\sim 10^{-2}\,\mathrm{km\, s^{-1}}$, where $c_{\mathrm{s}}$ is the sound speed in the disc \citep{krticka2011}. This value is much smaller than the outflow velocity of $5\,\mathrm{km\,s^{-1}}$ adopted by \citet{waters1987}.

% motivation
Observational evidence has strengthened our confidence in the VDD model, and there have been many IR missions since \textit{IRAS} and \citet{waters1987} presented their results. In light of these theoretical and observational advancements, it is now time to revisit the \citeauthor{waters1987} study, using more recent observations and a better theoretical formalism. In this work, we extend the pseudo-photosphere model to include the stellar rotation effects, and a better prescription for the stellar flux attenuated by the disc. Using this model, we investigate the disc properties of 80 Be stars, based on their IR spectral energy distribution (SED). The model improvements are described in Section~\ref{sect:pphot_model}. Next, we list the sample selection criteria (Section~\ref{sect:sample}), and present the SED fits (Section~\ref{sect:sed_fit}). Finally, we interpret our results using VDD hydrodynamical simulations (Section~\ref{sect:discussion}), and the conclusions follow.

% MODEL DESCRIPTION______________________________

\section{The pseudo-photosphere model}
\label{sect:pphot_model}

The pseudo-photosphere is defined as the region where $\tau\geq \overline{\tau}$, where $\tau$ is the total disc optical depth along the line of sight, and $\overline{\tau}$ is a free parameter close to one. The model assumes a geometrically thin isothermal disc at temperature $T_{\mathrm{d}}=f\,T_{\mathrm{eff}}$, where $f=0.6$ \citep{carciofi2006} and $T_{\mathrm{eff}}$ is the stellar effective temperature. The simplified parametric description of a VDD density profile is given by (e.g., \citealt{bjorkman2005}):
\begin{equation}
\rho(\varpi, z) = \rho_0\left(\frac{\varpi}{R_{\mathrm{eq}}}\right)^{-n}\exp\left(-\frac{z^2}{2H^2}\right),
\end{equation}
where $\rho_0$ is the disc base density, $R_{\mathrm{eq}}$ is the equatorial stellar radius, $\varpi$ and $z$ are respectively the radial and vertical cylindrical coordinates in the stellar frame of reference, $H\propto \varpi^{\beta}$ is the disc scale height, and $\beta$ the disc flaring exponent. \citet{vieira2015} derived a semi-analytic expression for the pseudo-photosphere size $\overline{R}$ as a function of the stellar and disc parameters,
\begin{equation}\label{eq:ref}
\overline{R}\propto [\rho_0^2\,\lambda^{2+u}]^{1/(2n-\beta)},
\end{equation}
where
\begin{equation}
u=\frac{d\ln(g_{\mathrm{ff}}+g_{\mathrm{bf}})}{d\ln\lambda},
\end{equation}
and $g_{\mathrm{ff}}$ and $g_{\mathrm{bf}}$ are the free-free and bound-free gaunt factors, respectively. The model calibration and validation were made based on \texttt{HDUST} results \citep{carciofi2006}. \texttt{HDUST} is a three-dimensional Monte Carlo radiative transfer (RT) code, capable of simultaneously solving the non-LTE hydrogen level populations, ionization fraction and electron temperature from the radiative equilibrium condition at each disc position. By adopting $\overline{\tau}=1.3$, \citet{vieira2015} reproduced the \texttt{HDUST} IR fluxes to within $10\%$. Finally, note the pseudo-photosphere model assumptions are only valid for disc inclinations $\lesssim 75^{\circ}$. Above this limit, the geometrically thin disc approximation no longer holds.

\subsection{Stellar rotation effects}
\label{sect:stellar_rot_effects}

Be stars are fast rotators (e.g., \citealt{porter1996}, \citealt{fremat2005}, \citealt{rivinius2006}). Although the stellar mass loss mechanism remains unknown, rotation close to the critical velocity \citep[$W\equiv v_{\mathrm{rot}}/v_{\mathrm{orb}}\gtrsim 0.7$,][]{rivinius2013} certainly represents an important ingredient for the Be phenomenon. Aside from its relevance to the stellar evolution (\citealt{ekstrom2008}, \citealt{georgy2013}), the fast rotation also causes flattening and gravity darkening of the star \citep{vonzeipel1924}. To take stellar oblateness into account in the pseudo-photosphere model, we employ the Roche approximation \citep[e.g.,][]{cranmer1996, ekstrom2008}. The geometrical deformation of the star modifies the stellar emitting area, and also determines the fraction of the disc emission blocked by the star.

\begin{figure}%[t]
\begin{center}
\includegraphics[width=1.\linewidth]{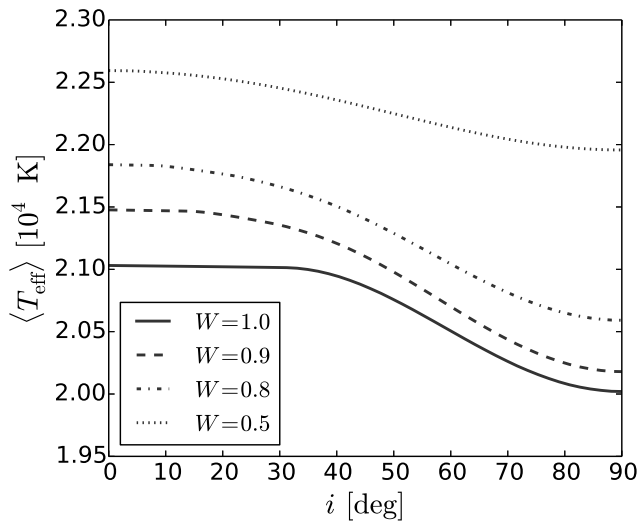}
\caption{$\langle T_{\mathrm{eff}}\rangle$ as a function of inclination. Each line represents a different stellar rotation rate, as indicated. The remaining stellar parameters ($M_{\star}$, $R_{\mathrm{pole}}$ and $\log L_{\star}$) correspond to a $B2$ sub-spectral type (Table~\ref{tab:stellar_par}). \label{fig:teff_avg}}
\end{center}
\end{figure}

Rather than taking into account the detailed latitude dependence of the stellar surface effective temperature, we instead use its average value over the stellar hemisphere facing the observer:
\begin{equation}\label{eq:teff_avg}
\langle T_{\mathrm{eff}}\rangle^4 = \frac{1}{A_{\star}}\int_{A_{\star}} T_{\mathrm{eff}}^4(\theta)\,dA,
\end{equation}
where $A_{\star}$ is the stellar surface projected in the plane of the sky, $\theta$ is the stellar co-latitude,
\begin{equation}
T_{\mathrm{eff}}(\theta) = \left(\frac{L_{\star}/\sigma_{\textrm{B}}}{\oint g^{4b}\,dA} \right)^{1/4} g^b(\theta),
\end{equation}
\citep{cranmer1996}, $L_{\star}$ the stellar luminosity, $\sigma_{\mathrm{B}}$ the Stefan-Boltzmann constant, and $b$ is the gravity darkening exponent. The integral in the denominator is computed over the entire stellar surface, and we adopt the prescription described by \citet{espinosa2011} to compute $b$ for a given rotation rate. This average approximation is adequate for our purposes, since the total stellar flux is an integrated quantity. Figure~\ref{fig:teff_avg} shows $\langle T_{\mathrm{eff}}\rangle$ as a function of stellar rotation rate and inclination for a $B2$ star (see Table~\ref{tab:stellar_par}).

%-------------------------------------------------
% Teff avg
%-------------------------------------------------
%-------------------------------------------------

%-------------------------------------------------
% Table stellar parameters
%-------------------------------------------------
\begin{table}%[!t]
\caption{List of adopted stellar parameters, derived from the interpolation of the evolutionary models computed by \citet{georgy2013}.}
\label{tab:stellar_par}
\begin{center}
\begin{tabular}{lcccc}
\hline
Parameter & \multicolumn{4}{c}{Value} \\ \hline
Sp. Type	&	$B1V$	&	$B2V$	&	$B3V$	&	$B7V$	\\
$M_{\star}/{\rm M_{\odot}}\,^a$	&	$12.5$	&	$9.6$	&	$7.7$	&	$4.2$	\\
$R_{\mathrm{pole}}/{\rm R_{\odot}}$	&	$6.7$	&	$5.6$	&	$4.9$	&	$3.7$	\\
%$L_{\star}/{\rm L_{\odot}}$	&	$18550$	&	$7830$	&	$3750$	&	$450$	\\
$\log(L_{\star}/{\rm L_{\odot}})$	&	$4.3$	&	$3.9$	&	$3.6$	&	$2.7$	\\
%$T_{\textrm{eff}}/\mathrm{K}$	&	$24100$	&	$21200$	&	$18760$	&	$12750$	\\
$\langle T_{\textrm{eff}}\rangle_{\mathrm{pole}} /\mathrm{K}$	&	$24\, 800$	&	$21\, 800$	&	$19\, 300$	&	$13\, 150$	\\
$\langle T_{\textrm{eff}}\rangle_{\mathrm{eq}} /\mathrm{K}$	&	$23\, 400$	&	$20\, 600$	&	$18\, 200$	&	$12\, 400$	\\
$W\,^b$	&	\multicolumn{4}{c}{$0.8$}	\\
$t/t_{\textrm{MS}}$	&	\multicolumn{4}{c}{$0.75$}	\\
\hline
\end{tabular}
\end{center}
{\it Note.} $^{a}$\citet{townsend2004}, $^{b}$ mean value for Be stars \citep{rivinius2013}.
\end{table}
%-------------------------------------------------

%-------------------------------------------------
% synthetic image
%-------------------------------------------------
\begin{figure}%[t]
\begin{center}
\includegraphics[angle=0,scale=.5]{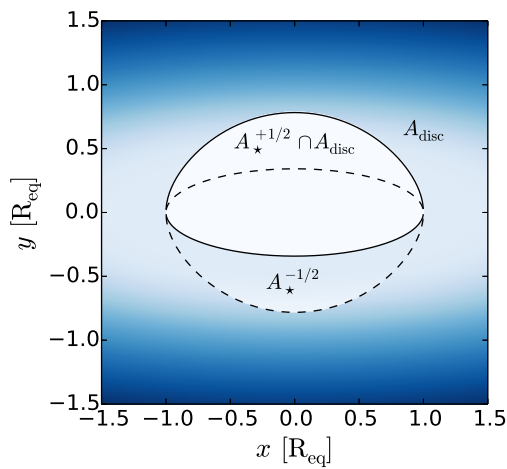}
\caption{Synthetic image of a rotation-flattened star, surrounded by a gaseous disc. The solid line represents the contour of the stellar upper hemisphere ($A_{\star}^{+1/2}$), while the dashed contours represent the hidden stellar boundaries. The stellar lower hemisphere flux is attenuated by the disc opacity. The adopted parameters are $W=0.8$, $i=70^{\circ}$, $\overline{R}/R_{\mathrm{eq}}=2$ and $n=3$. The disc size was assumed to be much larger than $R_{\mathrm{eq}}$. \label{fig:synthetic_img}}
\end{center}
\end{figure}
%-------------------------------------------------

\subsection{Stellar flux attenuation}
\label{sect:attenuation}

%-------------------------------------------------
% hdust excess
%-------------------------------------------------
%\afterpage{
\begin{figure*}%[t]
\begin{center}
\includegraphics[width=0.7\linewidth]{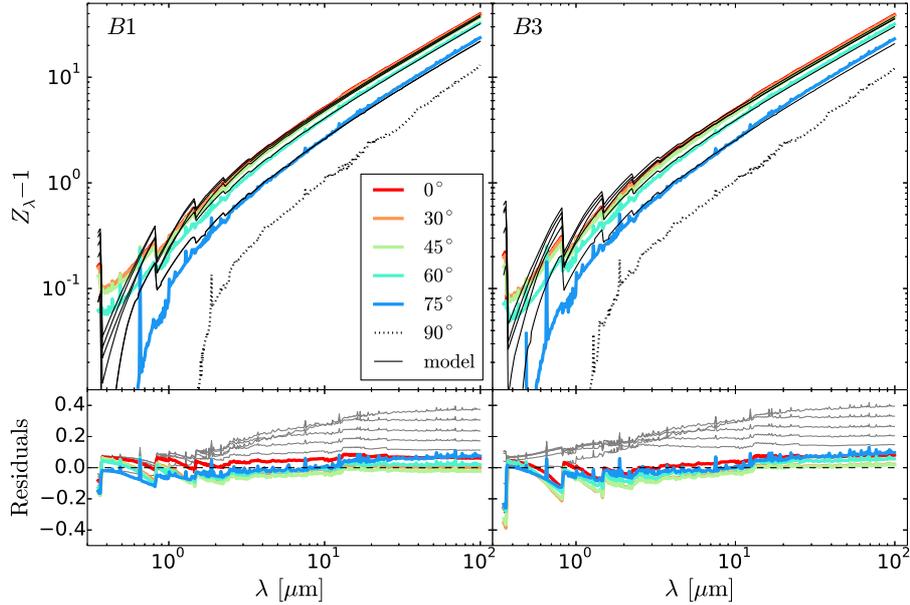}
\caption{Comparison between the pseudo-photosphere model (thin black lines) and \texttt{HDUST} results (thick coloured lines), for a disc with $\rho_0=8.4\times 10^{-12}\,\mathrm{g\, cm^{-3}}$ and $n=3.5$. Upper panels: flux excesses computed with the pseudo-photosphere model (stellar rotation and attenuation effects included) superimposed on the \texttt{HDUST} results. Bottom panels: residual plots, comparing the \texttt{HDUST} results to the pseudo-photosphere model including (i) the stellar rotation effects (thick lines; colors representing disc inclination), and (ii) neglecting those effects (thin grey lines; the largest value corresponding to model with $i=75^{\circ}$). The edge-on \texttt{HDUST} model (dotted line) is presented in the upper panels just for reference, since it does not have a pseudo-photosphere model counterpart. The adopted stellar parameters correspond to a $B1$ (left) and a $B3$-type (right), and are presented in Table~\ref{tab:stellar_par}. \label{fig:hdust_excess}}
\end{center}
\end{figure*}
%}
%-------------------------------------------------

The pseudo-photosphere model \citep{vieira2015} has three possible cases: (i) the general case, where both the pseudo-photosphere and tenuous region are present, (ii) the tenuous case, where the disc is entirely optically thin, and (iii) the case where the pseudo-photosphere is truncated, which means that $\overline{R}$ exceeds the disc size. Different flux expressions were derived for each one of these cases. For simplicity, the detailed stellar extinction caused by the disc was neglected, and the stellar flux contribution either arises from both stellar hemispheres in the tenuous case, or only from the hemisphere above the stellar equator when the pseudo-photosphere is present. However, such approximation causes a spectral energy distribution (SED) slope discontinuity at $\lambda'$ such that $\overline{R}(\lambda')/R_{\mathrm{eq}}=1$, which is more evident for higher inclinations. In order to remove this non-physical artifact, the pseudo-photosphere model flux expressions were generalized to properly include the stellar flux attenuation caused by the disc. Consider the flux excess definition:
\begin{equation}
Z_{\lambda} = \frac{F_{\lambda}}{F_{\lambda}^{\star}},
\end{equation}
where $F_{\lambda}$ is the total flux (star and disc combined),
\begin{equation}
F_{\lambda}^{\star} = \frac{A_{\star}}{d^2}S_{\lambda}^{\star}(\langle T_{\mathrm{eff}}\rangle)
\end{equation}
is the flux of the disc-less star, and $S_{\lambda}^{\star}$ is the stellar surface brightness. To compute the stellar flux, we adopted interpolated models from \citet{castelli2003}. The stellar brightness is assumed to be uniform (no limb darkening), and is a function of $\langle T_{\mathrm{eff}}\rangle$. The new flux excess expression for the general case may be written as
\begin{equation}
Z_{\lambda} = \frac{\cos i\, \mathscr{F}}{A_{\star}/(\pi R_e^2)}\,\left[\left(\overline{R}/R_{\mathrm{eq}}\right)^2\Psi - 1 \right] + Z_{\lambda}^{\mathrm{num}},
\end{equation}
where
\begin{align}
Z_{\lambda}^{\mathrm{num}} &= \frac{A_{\star} -A_{\star}^{-1/2}}{A_{\star}} + \frac{1}{A_{\star}} \int_{A_{\star}^{-1/2}} e^{-\tau}\,dA \nonumber\\
 & \quad - \frac{1}{A_{\star}}\int_{A_{\star}^{+1/2} \cap A_{\mathrm{disc}}} \mathscr{F}\,(1-e^{-\tau})\, dA
\end{align}
is the numerically computed flux excess component (see Figure~\ref{fig:synthetic_img} for the definitions of the integration domains),
\begin{equation}
\Psi = 1 + \frac{2\overline{\tau}}{2n-\beta-2}\left[1-(R_{\mathrm{d}}/\overline{R})^{-2n+\beta+2}\right],
\end{equation}
 $\mathscr{F} \equiv S_{\lambda}^{\star} / B_{\lambda}(T_{\mathrm{d}})$, $R_{\mathrm{d}}$ is the disc size, and $\tau$ follows the definition given by \citet{vieira2015}. Because of the complicated Roche model geometry, the integrals can only be numerically evaluated (except for the trivial pole-on case). The derived expression now accounts for the attenuated stellar flux, and properly subtracts the disc flux contribution shadowed by the star. The expressions for the entirely tenuous disc and truncated pseudo-photosphere become, respectively,
\begin{align}
Z_{\lambda}^{\mathrm{tenuous}} &= \frac{2\overline{\tau}\cos i \mathscr{F}}{A_{\star}/(\pi R_{\mathrm{eq}}^2)}\left[ \frac{1 - (R_{\mathrm{d}}/R_{\mathrm{eq}})^{-2n+\beta+2}}{2n-\beta-2}\right] \left(\overline{R}/R_{\mathrm{eq}}\right)^{2n-\beta} \nonumber\\
 & \quad + Z_{\lambda}^{\mathrm{num}}
\end{align}
and
\begin{align}
Z_{\lambda}^{\mathrm{trunc}} &= \frac{\cos i\, \mathscr{F}}{A_{\star}/(\pi R_{\mathrm{eq}}^2)}\,\left[\left(R_{\mathrm{d}}/R_{\mathrm{eq}}\right)^2 - 1\right] + Z_{\lambda}^{\mathrm{num}}.
\end{align}
The comparison between the newly derived formulae and \texttt{HDUST} results is presented in Figure~\ref{fig:hdust_excess}. Like the previous version, the derived IR fluxes are accurate within $10\%$ when compared to \texttt{HDUST} results. In the same figure, the bottom panels also show the residuals for the pseudo-photosphere model without the rotation effects. For this case, the residuals increase with inclination. At larger inclinations, the disc area hidden by the non-rotating star is larger than the one hidden by a flattened rotating star. Consequently, an important contribution of the disc flux is missed when rotation is neglected. The dependence of the magnitude excess as a function of the disc base density is presented in Figure~\ref{fig:deltmag}.

% SAMPLE SELECTION________________________________

\section{Sample selection}
\label{sect:sample}

\citet{fremat2005} determined the fundamental parameters of 130 early-type stars based on their observed spectra, taking into account the stellar rotation effects. To select our sample of Be stars from their list, the following criteria were applied:

\begin{itemize}[leftmargin=.2in]
\item[a)] classical Be star classification;
\item[b)] non-shell line profile designation;
\item[d)] having at least two IR flux bands measured by the same mission (i.e., at similar epochs);
\item[e)] observed IR spectral slope $d\ln F_{\lambda}/d\ln\lambda$ lying between $-4$ and $-1.5$;
\item[f)] observed flux larger than the model stellar flux.
\end{itemize}

The IR flux data from the \textit{IRAS} \citep{neugebauer1984}, \textit{AKARI/IRC} \citep{ishihara2010}, and \textit{AllWISE} \citep{wright2010} missions were used. Of the \textit{AllWISE} fluxes, only those for $12$ and $22\,\mathrm{{\mu}m}$ were used because the pseudo-photosphere model predictions are less reliable at $\lambda\lesssim 5\,\mathrm{{\mu}m}$ \citep{vieira2015}. Excluding the \textit{AllWISE} shorter wavelength measurements has the additional advantage that the three adopted IR missions then probe similar regions of the disc (\textit{IRAS} provides fluxes at $12$, $25$ and $60\,\mathrm{{\mu}m}$ [we excluded $100\,\mathrm{{\mu}m}$], and \textit{AKARI/IRC} at $9$ and $18\,\mathrm{{\mu}m}$). Finally, upper limit measurements were also excluded from our analysis.

Special care was given to eliminate shell stars from the list of \citet{fremat2005}, because the model is valid only for $i\lesssim 75^{\circ}$. This was done by visually inspecting the spectra available in online repositories, such as \textit{BeSS}\,\footnote{\url{http://basebe.obspm.fr/basebe/}}, to exclude the objects with clear shell signatures. The objects whose inclination angle (as determined by \citeauthor{fremat2005}) that were larger than $75^{\circ}$ without clear shell features in the spectrum were kept in our list. For these, the inclination angle was set to $i=75^{\circ}\pm 10^{\circ}$. Criterion (e) ensures a disc is present at the time of the observations, and rules out flat/increasing slopes, which may be indicative of the presence of dust. Criterion (f) eliminates objects with inconsistent stellar parameters. After applying the above criteria, 80 out of the original 130 stars were selected. Their fluxes were color-corrected by computing the spectral slope from the catalogue values as a first guess (separately for each mission), and then using the mission bandpasses to iterate the monochromatic fluxes until convergence is achieved. The resulting fluxes are presented in Appendix~\ref{appendix:sample}.

% SED fitting______________________________

\section{SED fitting}
\label{sect:sed_fit}

%-------------------------------------------------
% delta magnitude
%-------------------------------------------------
\begin{figure}%[t]
\begin{center}
\includegraphics[width=1.\linewidth]{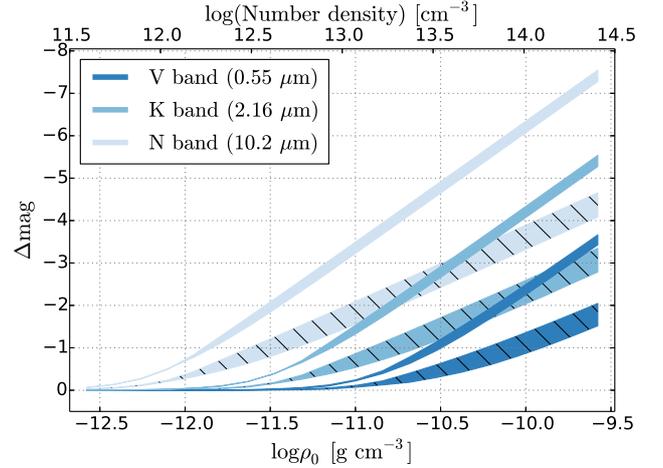}
\caption{Disc flux excess (magnitudes) as a function of the disc base density for a $B2$ star. The curve width covers disc inclinations from $0^{\circ}$ to $75^{\circ}$ (the pole-on case corresponds to the $\Delta\mathrm{mag}$ upper limit). The wavelength is specified by the blue shade, as indicated. The hatched curves correspond to an $n=3.5$ model, while the non-hatched curves represent the $n=2.5$ models. \label{fig:deltmag}}
\end{center}
\end{figure}
%-------------------------------------------------

The observed SEDs were fitted using the \texttt{emcee} code \citep{foreman2013}, a Markov chain Monte Carlo (MCMC) implementation. \texttt{emcee} samples the posterior probability in an $n$-dimensional parameter space, given a likelihood function ($\propto \exp[-\chi^2/2]$ in our case). For each simulation, we used $100$ walkers (random-walk samplers) with $100$ steps in the initial phase (burn-in) and $1000$ steps in the final sampling phase (starting from the last state of the burn-in chain). The following parameters of the pseudo-photosphere model were kept fixed: $f=0.6$, $\overline{\tau} = 1.3$, $\beta = 1.5$ and $R_{\mathrm{d}} = 1000\,R_{\mathrm{eq}}$. The stellar parameters from \citeauthor{fremat2005} (\citeyear{fremat2005}; $T_{\mathrm{eff,\, 0}}$, $\log g_0$, $v\sin i$ and $V_{\mathrm{c}}$)\,\footnote{The ``0'' subscripts refer to the parent non-rotating counterpart parameters (pnrc), defined by \citet{fremat2005}.}, and the Hipparcos parallaxes were chosen to vary within their 1-$\sigma$ confidence interval in the MCMC run. This procedure ensures that the uncertainties in the stellar parameters are properly accounted for when estimating the confidence intervals of the disc parameters. Appendix~\ref{appendix:stellar_par} describes how the stellar parameters of interest were estimated based on the parameters derived by \citet{fremat2005}.

When available, interferometric measurements were used to estimate the disc inclination (Table~\ref{tab:interf_incl}). Otherwise, a $10^{\circ}$ confidence interval was adopted for the inclination values from \citet{fremat2005}, since their original confidence ranges were probably underestimated \citep[][see also the typical values of Table~\ref{tab:interf_incl}]{rivinius2013}. Finally, no prior constraints were applied to the parameters of interest, $n$ and $\log\rho_0$, except for restricting $n$ to positive values. A total of $169$ models were fitted with \texttt{emcee}, since many of the $80$ objects were observed by different missions.

%-------------------------------------------------
% Table interferometric inclinations
%-------------------------------------------------
\begin{table}
\begin{center}
\caption{Disc inclinations derived from interferometric measurements. For the case of axial ratio measurements, we adopted the geometrically thin disc approximation $i = \cos^{-1}(\mathrm{axial\, ratio})$. \label{tab:interf_incl}}
\begin{tabular}{lcl}
\hline
HD & inclination & Reference\\ \hline
5394 & $39.6^{\circ}\pm 1.8^{\circ}$ & \citet{quirrenbach1997}\\
23630 & $41.1^{\circ}\pm 7.0^{\circ}$ & \citet{tycner2005}\\
25940 & $40.4^{\circ}\pm 19.7^{\circ}$ & \citet{delaa2011}\\
37795 & $35^{\circ}\pm 5^{\circ}$ & \citet{meilland2012}\\
50013 & $35^{\circ}\pm 10^{\circ}$ & \citet{meilland2012}\\
58715 & $46.4^{\circ}\pm 16.4^{\circ}$ & \citet{tycner2005}\\
89080 & $65^{\circ}\pm 10^{\circ}$ & \citet{meilland2012}\\
91465 & $70^{\circ}\pm 10^{\circ}$ & \citet{meilland2012}\\
105435 & $35^{\circ}\pm 15^{\circ}$ & \citet{meilland2012}\\
120324 & $25^{\circ}\pm 5^{\circ}$ & \citet{meilland2012}\\
158427 & $45^{\circ}\pm 5^{\circ}$ & \citet{meilland2012}\\
217891 & $45.4^{\circ}\pm 12.1^{\circ}$ & \citet{touhami2013}\\
\hline
\end{tabular}
\end{center}
\end{table}
%-------------------------------------------------

\subsection{Results}
\label{sect:results}

The SED fitting results are presented in Appendix~\ref{appendix:results} (Table~\ref{tab:results}), where the median values of the derived posterior probability distributions are given. The derived uncertainties correspond to the 16$^{\mathrm{th}}$ and 84$^{^{\mathrm{th}}}$ percentiles of these distributions, which are equivalent to a 1-$\sigma$ Gaussian variance. The associated $\overline{R}$ values, as well as a discussion about the effects of possible  disc truncation effects caused by a binary companion are discussed in Appendix~\ref{appendix:ref_trunc}. Table~\ref{tab:results} also lists the values for the steady-state decretion rate, defined as
\begin{equation}\label{eq:mss}
\dot{M}_{\mathrm{SS}}=\frac{3\pi\sqrt{2\pi}\alpha\,R_{\mathrm{eq}}^2\, c_{\mathrm{s}}^3\,\rho_0}{V_{\textrm{crit}}^{2} \left[\left(R_0/R_{\mathrm{eq}}\right)^{1/2}-1\right]},
\end{equation}
where $\alpha$ is the viscosity parameter \citep{shakura1973}, $c_{\mathrm{s}}=(kT_{\textrm{d}}/\mu m_{\textrm{H}})^{1/2}$ is the isothermal sound velocity, $V_{\textrm{crit}}=\sqrt{G M_{\star}/R_{\mathrm{eq}}}$ is the break-up velocity, $M_{\star}$ is the stellar mass, $T_{\textrm{d}}$ is the disc temperature, $\mu$ is the mean molecular weight of the gas, $m_H$ is the atomic mass unity and $k$ is the Boltzmann constant. The integration constant $R_0$ is approximated by the isothermal critical radius \citep{krticka2011,okazaki2001}:
\begin{equation}
\frac{R_0}{R_{\mathrm{eq}}} = \frac{R_{\textrm{c}}}{R_{\mathrm{eq}}}=\frac{3}{10}\left(\frac{V_{\textrm{crit}}}{c_{\mathrm{s}}}\right)^2.
\end{equation}
We emphasize however that Equation~(\ref{eq:mss}) is strictly true only for a disc fed long enough to approach a steady-state. The results from Eq.~(\ref{eq:mss}) should be regarded as estimates of the mass decretion rate, not true determinations.

\begin{figure*}%[!ht]
\begin{center}
\subfigure[Sampling of $\beta$~CMi disc parameters.]{%
\label{fig:posterior_betcmi}
\includegraphics[width=.6\linewidth]{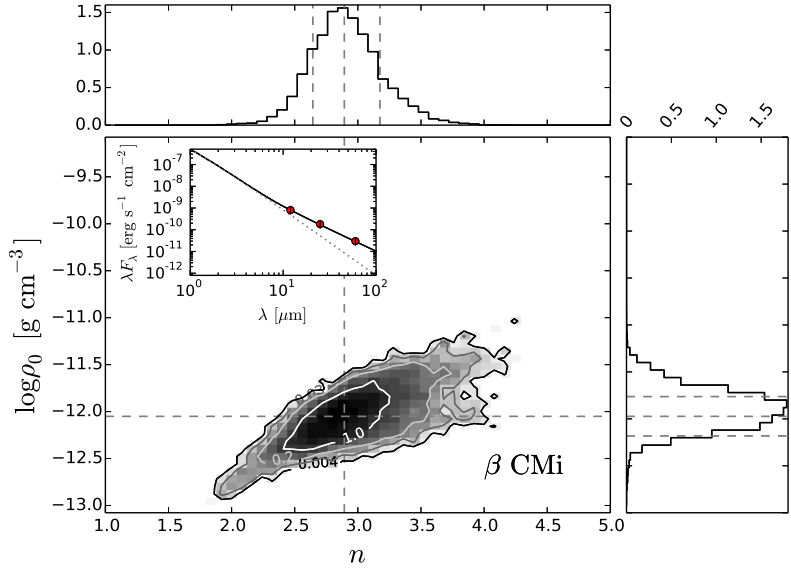}
}\\
\subfigure[Sampling of 28~Cyg disc parameters.]{%
\label{fig:posterior_28cyg}
\includegraphics[width=.6\linewidth]{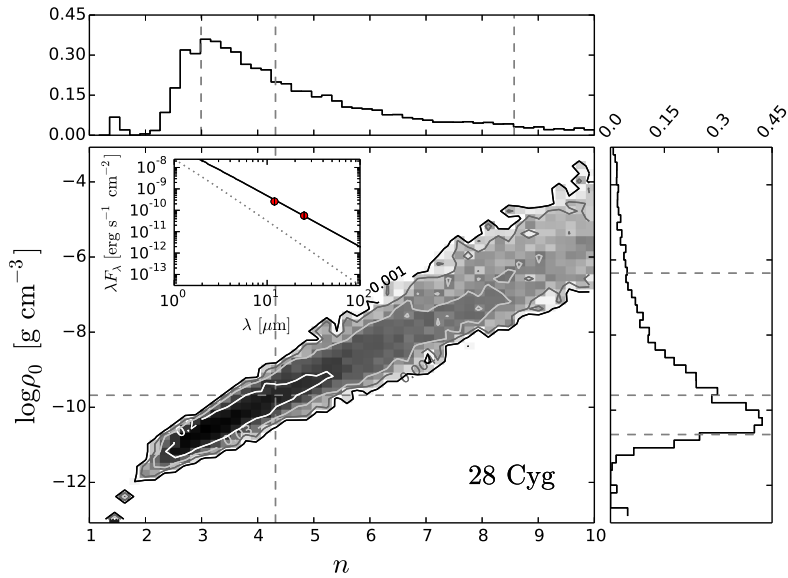}
}\\
\end{center}
\caption{Examples of the posterior probability distributions of disc parameters computed with the \texttt{emcee} code. The contour values indicate the probability density levels, with its integral normalized to unity. The SED fit is shown inset within the main panel, where the circles correspond to \textit{IRAS} observations, the solid line represents the total flux, and the dotted line represents the photospheric flux. The dashed lines in the main panel indicate the $50^{\mathrm{th}}$ percentiles of the distributions (i.e., its median value). The upper and right panels show the probability densities for the individual parameters $\log\rho_0$ and $n$, respectively. The dashed lines correspond to the $16^{\mathrm{th}}$, $50^{\mathrm{th}}$, and $84^{\mathrm{th}}$ percentiles of their respective distributions (see text). Notice that panels (a) and (b) have different plot ranges.}
\label{fig:sampling_example}
\end{figure*}

\begin{figure}
\begin{center}
\includegraphics[width=.99\linewidth]{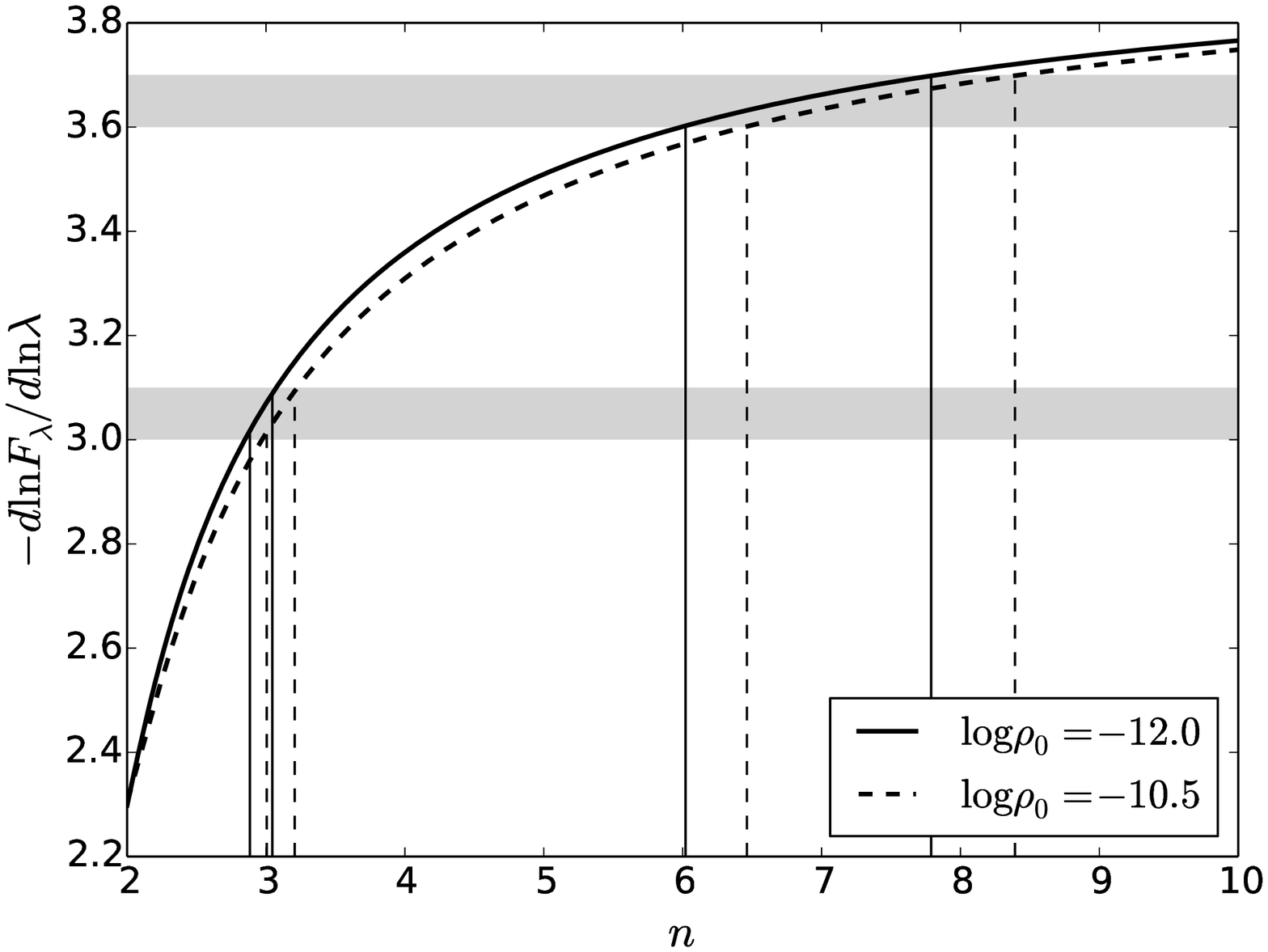}
\caption{Spectral slope between $10\,\mathrm{\mu m}$ and $20\,\mathrm{\mu m}$ as a function of $n$, for two disc base densities. The horizontal grey bars indicate a confidence range of $0.1$ (a typical value found for the observed SED slopes) at two arbitrary slope values. The vertical lines indicate the corresponding $n$ confidence ranges. For both curves, we adopted a \textit{B2} spectral type (Table~\ref{tab:stellar_par}) and a pole-on orientation. \label{fig:n_slp}}
\end{center}
\end{figure}

Figure~\ref{fig:sampling_example} shows the \texttt{emcee} sampling results for $\beta$~CMi and 28~Cyg, as representative cases. In particular, the results found for $\beta$~CMi are in very good agreement with those found by \citet{klement2015}, who found that the observations probing a more extended region of the disc are compatible with $n=3$ and $\rho_0=2\times 10^{-12}\,\mathrm{g\,cm^{-3}}$. Note there is a correlation between $n$ and $\log\rho_0$, since different combinations of these parameters can result in a similar $\overline{R}$ (see Equation~\ref{eq:ref}). Conversely, the derived uncertainties for these parameters are also correlated (Table~\ref{tab:results}). Additionally, the results show that large $n$ values ($n\gtrsim 3.5$, case of 28~Cyg) usually have broader confidence ranges (and, consequently, larger $\log\rho_0$ uncertainties). To understand this behaviour, one must recall that the spectral slope is mainly determined by $n$ \citep{vieira2015}, and has a weak dependence on $\log\rho_0$ and disc inclination. Figure~\ref{fig:n_slp} shows this dependence on $n$ for two base densities. For a fixed uncertainty in the SED slope, the uncertainties for a large $n$ are much larger than those for a smaller $n$ value.

\section{Discussion}
\label{sect:discussion}

%-------------------------------------------------
% sampling example
%-------------------------------------------------
%-------------------------------------------------

%-------------------------------------------------
% n vs. spectral slope
%-------------------------------------------------
%-------------------------------------------------

\begin{figure*}[!b]
\begin{center}
\includegraphics[angle=0,scale=.6]{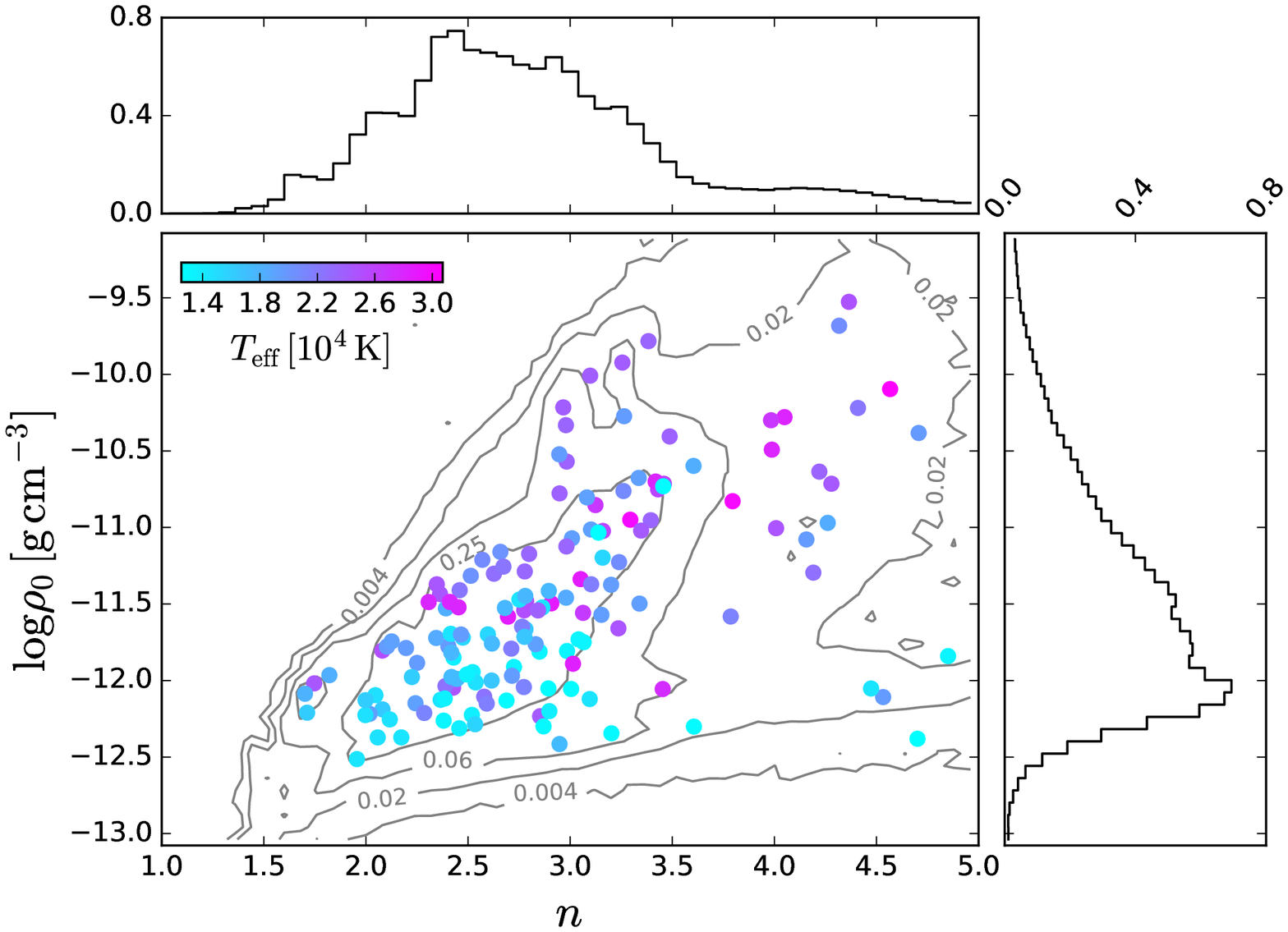}
\caption{Disc density and slope of the sample of Be stars. Shown are the \texttt{emcee} results for the $n$ and $\log\rho_0$ distributions. The main panel shows the probability density (grey scale contours), while the upper and left panels show the distribution for the individual parameters. These distributions correspond to the combination of all posterior probabilities derived with the \texttt{emcee} code, and their integral values were normalized to unity. The contour values correspond to the probability density levels. Superimposed, we plotted the median values of the \texttt{emcee} sampled distributions for the individual stars. The colors indicate the stellar effective temperature, computed by \citet{fremat2005}. \label{fig:n_ln0_histo2d}}
\end{center}
\end{figure*}

Figure~\ref{fig:n_ln0_histo2d} shows the disc parameters of the selected sample. The grey shading in the diagram was obtained by combining the posterior probabilities of $n$ and $\rho_0$ for all $169$ fitted SEDs, and normalizing its integral over the plane to unity. The median values of $n$ and $\log\rho_0$ of individual stars are shown as the colored circles, color-coded to indicate the effective temperature of the star.

Note the correlation between $n$ and $\log\rho_0$ along the high probability ridge of the shading plot. This cannot be due to the correlation shown in the individual fits, since the uncertainties for $n<5$ solutions lie typically between $0.1$ and $0.5$ for both parameters. The scatter distribution of $n$ vs. $\log\rho_0$ has a Pearson's correlation coefficient of $0.57$. There is a clear peak at $\log\rho_0\simeq -12$, with the base density spreading over two orders of magnitude, while the $n$ values preferentially occur between $1.5$ and $3.5$. Note the upper-left corner is practically empty.

\begin{figure*}%[!h]
\begin{center}
\subfigure[$\rho_0$ as \textit{vs.} $T_{\mathrm{eff}}$. The colors indicate the respective $n$ value, and the dashed line indicate the upper limit of the scatter distribution.]{%
\label{fig:teff_logrho0}
\includegraphics[width=.45\linewidth]{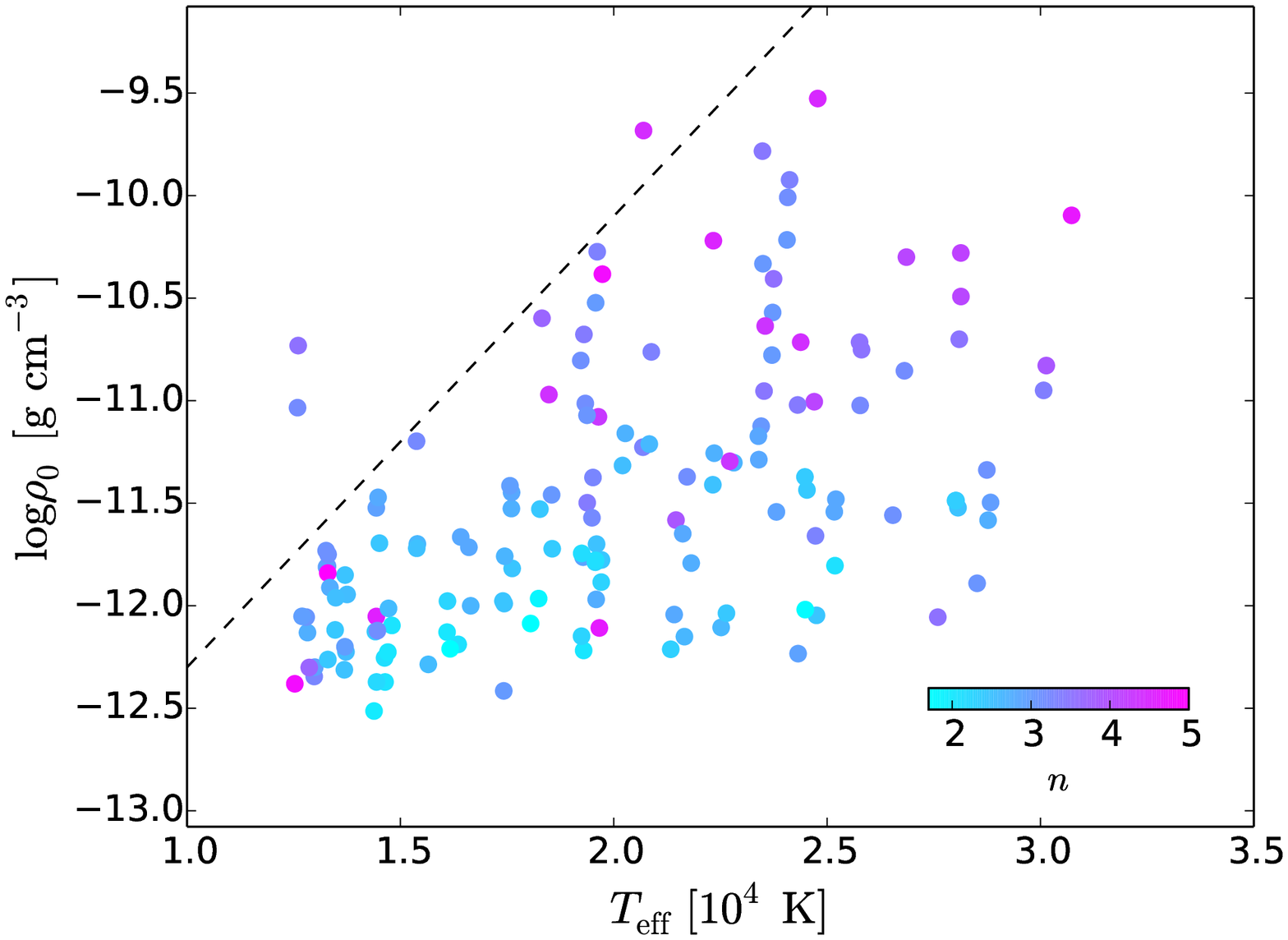}
}%
\subfigure[$n$ \textit{vs.} $T_{\mathrm{eff}}$. The colors indicate the respective $\log\rho_0$ values.]{%
\label{fig:teff_n}
\includegraphics[width=.45\linewidth]{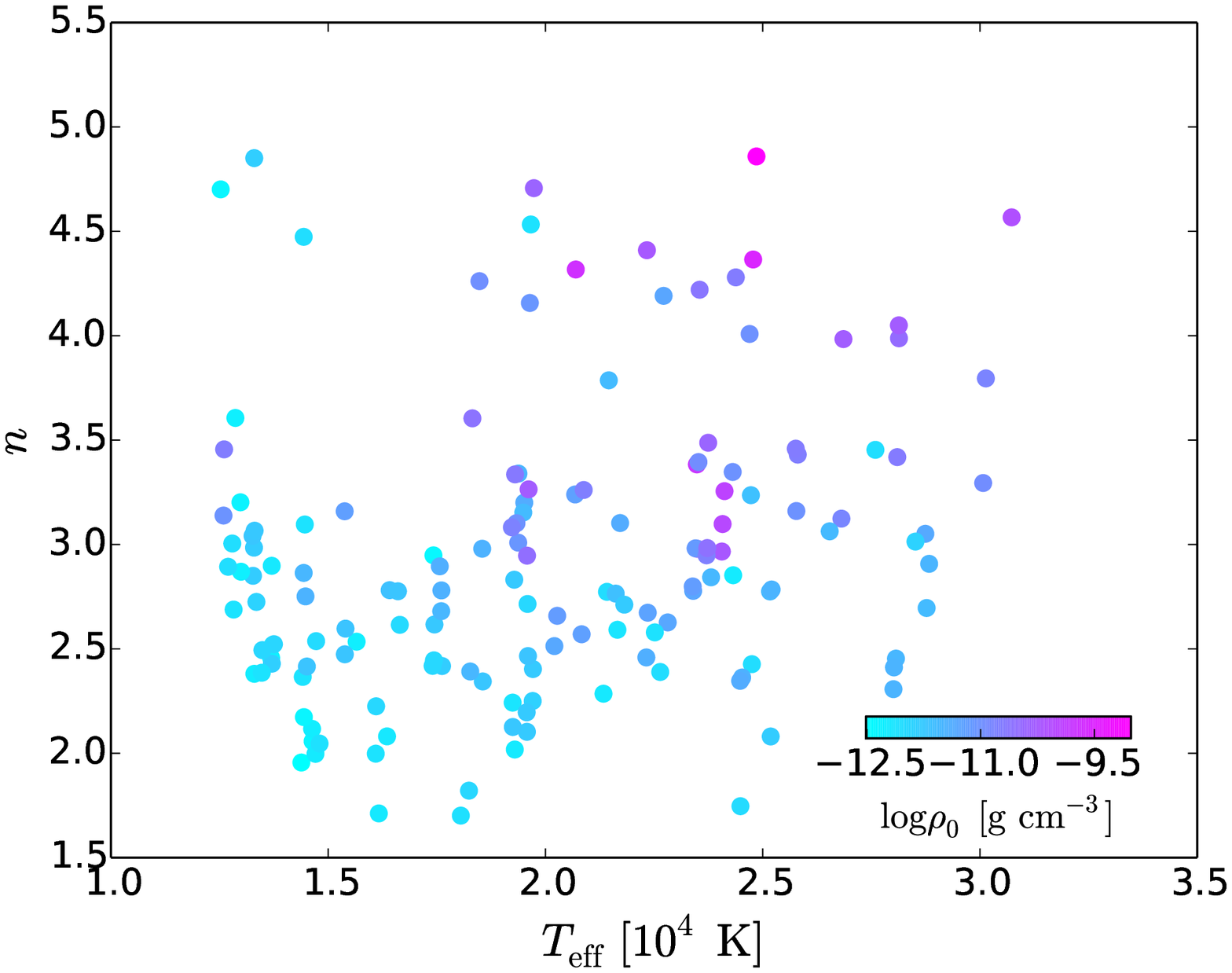}
}\\
\end{center}
\caption{Fitted parameters as a function of the stellar effective temperature. The models with $n>5$ were excluded.}
\label{fig:teff_par}
\end{figure*}

Another important property of Figure~\ref{fig:n_ln0_histo2d} is that earlier-type stars are more likely to have denser discs, while cooler stars usually occupy the lower region of the diagram. This result can better be seen in Figure~\ref{fig:teff_logrho0}, where stars for all temperatures can reach about the same low values, but only hot stars can reach higher ones. On the other hand, Figure~\ref{fig:teff_n} shows no clear correlation between $n$ and $T_{\mathrm{eff}}$.

%-------------------------------------------------
% n-logrho0 diagram
%-------------------------------------------------
%-------------------------------------------------

%-------------------------------------------------
% n/logrho0 vs Teff
%-------------------------------------------------
%-------------------------------------------------

\begin{figure}%[t]
\begin{center}
\includegraphics[width=1.\linewidth]{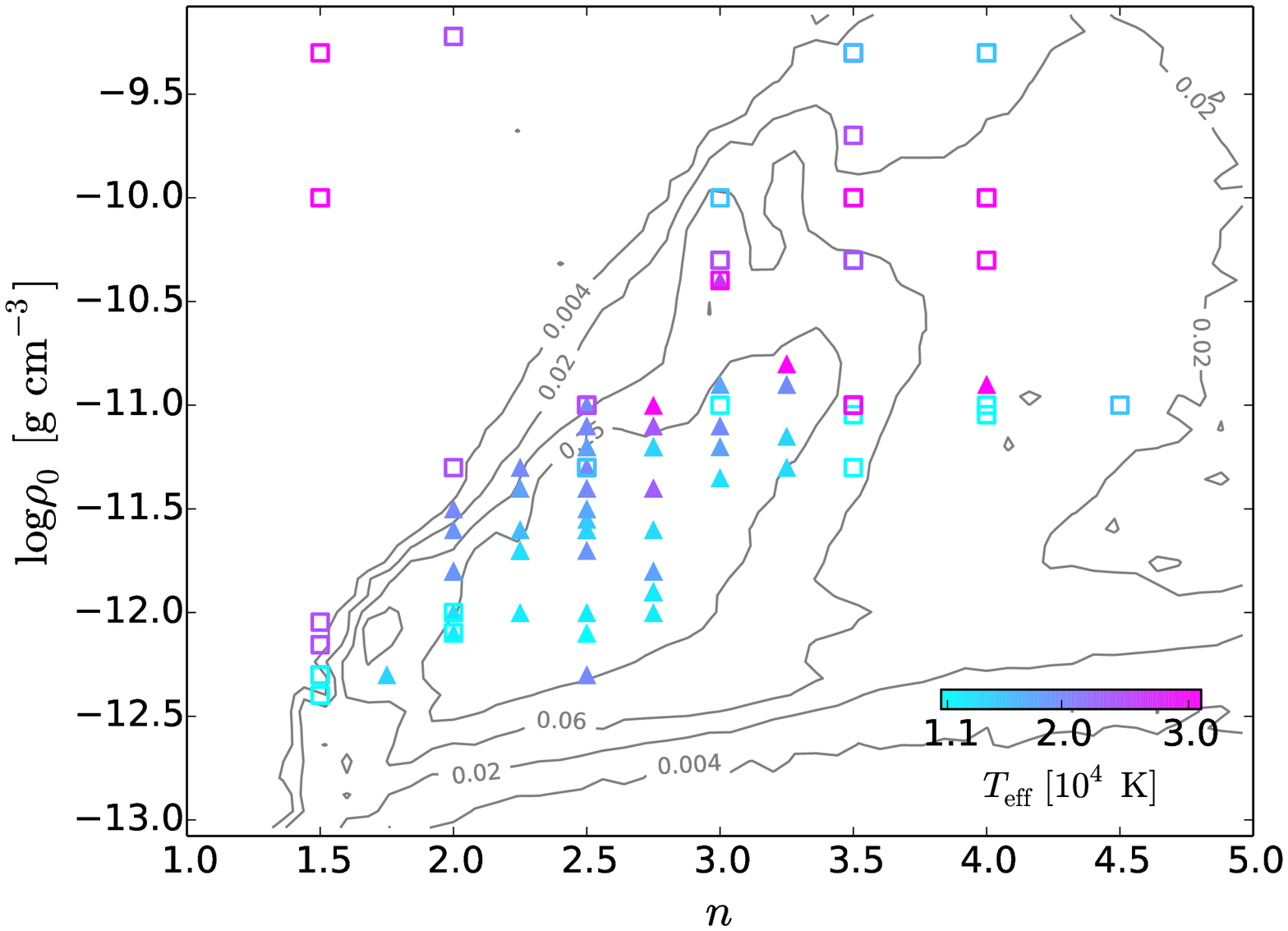}
\caption{Previous results for $n$ and $\log\rho_0$ found by \citet[][solid triangles]{waters1987} and \citet[][open squares]{silaj2010}, superimposed on the contours of the probability density distribution found by the present work. The symbol colors correspond to the stellar effective temperature adopted by each work. \label{fig:comparison}}
\end{center}
\end{figure}

Figure~\ref{fig:comparison} superimposes the previous results obtained by other authors on our distribution in the $n-\log\rho_0$ diagram. The choice of a disc flaring exponent of $\beta=1$ made by \citet{waters1987} is expected to result in values of $n$ smaller than ours by a difference of $0.25$. This occurs because the actual free parameter in both model formulations is $2n-\beta$ rather than $n$ alone, as discussed by \citet{vieira2015}. If we consider the sub-sample of \textit{IRAS} observations studied by \citet{waters1987}, the difference between our values of $n$ and the ones from those authors indeed occurs more often around $0.25$, although the differences range between $-0.4$ and $2.4$. Possible reasons for these differences are: \textit{(i)} our error estimates for $n$ are typically about $0.2$, i.e., of the same order that the expected differences; and \textit{(ii)} the inclusion of rotation effects, not taken into account by \citet{waters1987}, affects the SED shape (see Fig.~\ref{fig:hdust_excess}), and consequently the derived disc parameters. Despite of these effects, the general trend of the results found by \citet{waters1987} shows a good agreement with our results. They occupy the higher probability ridge found in this work, and present the same trend with $T_{\mathrm{eff}}$. However, the values obtained by \citet{silaj2010} from the fit of H$\alpha$ profiles appear to be systematically denser than our models. Such a difference may be related to a selection bias in their sample, which consists of Be stars with strong H$\alpha$ emission, and hence denser discs. Furthermore, \citeauthor{silaj2010} states that only preliminary estimates could be given based on their not extensive adopted grid.

\begin{figure}%[t]
\begin{center}
\includegraphics[width=1.\linewidth]{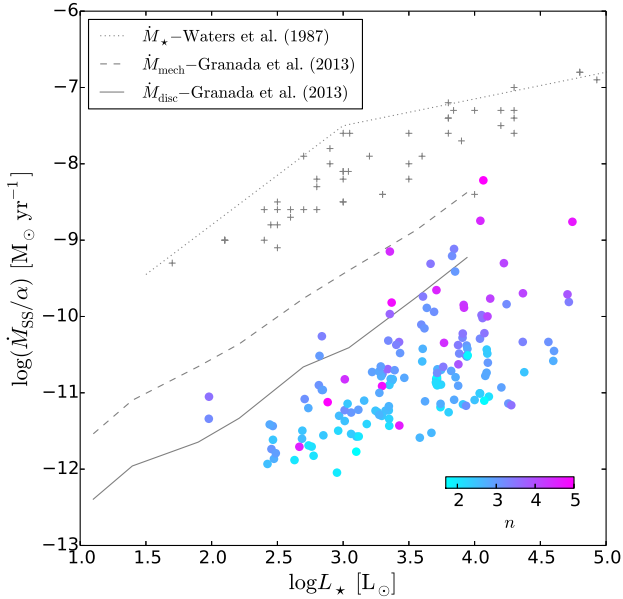}
\caption{Mass decretion rate as a function of stellar luminosity. The circles represent the values estimated from the \texttt{emcee} results, while the crosses indicate the values computed by \citet{waters1987}. The circle colors indicate the fitted $n$ value, and the $n>5$ cases were excluded. The dotted line corresponds to the upper limit suggested by \citet{waters1987}. The dashed and solid lines correspond, respectively, to the mechanical loss rate and the mass decretion rate computed by \citet{granada2013}. \label{fig:lum_mdot}}
\end{center}
\end{figure}

Figure~\ref{fig:lum_mdot} shows the relation between the stellar luminosity and the mass decretion rate. The estimated $\dot{M}_{\textrm{SS}}$ values tend to be higher for more luminous objects. This may indicate that more massive stars can provide higher mass injection rates. But another possibility is that discs of more massive stars have smaller $\alpha$ values, thereby making the radial diffusion time scale and outflow velocities smaller. As found before by \citet{vieira2015} for a smaller sample, our results for the mass decretion rates are up to three orders of magnitude smaller than the mass loss rates computed by \citet{waters1987}, and also show a larger scatter than that found by these authors. Additionally, we do not find the regime transition (i.e., a slope change in the dotted line) at $\log(L_{\star}/\mathrm{L_{\odot}})=3$ suggested by \citeauthor{waters1987}. Consequently, our results do not suggest different ejection mechanisms for early- and late-type Be stars, as proposed by them. Interestingly, our $\dot{M}_{\textrm{SS}}$ estimates are compatible with the $\dot{M}_{\mathrm{disc}}$ values calculated by \citet{granada2013}, which were based on a completely different approach. These authors proposed that the mechanical mass loss during the main-sequence evolution is that necessary to remove the angular momentum excess from an over-critically rotating stellar surface. From that, they estimated the disc structure and mass decretion rates using the model from \citet{krticka2011}.

\subsection{Disc variability}
\label{sect:variability}

Since the observations of each IR mission were taken in different epochs\,\footnote{Mission epochs: \textit{IRAS}: from January to November 1983; \textit{AKARI/IRC}: from May $2006$ to August $2007$; and \textit{AllWISE}: from January to November $2010$.}, disc variability can also be studied. The variation of the disc parameters, based on \textit{AKARI} and \textit{AllWISE} measurements, is presented in Figure~\ref{fig:evol_akari_wise}. The two more recent missions were selected because they have a smaller time separation ($\sim$$4$~yr) and more accurate fluxes. Interestingly, there are many more arrows moving from the upper right corner to the lower left one (i.e., from high $\log\rho_0$ and $n$) than the reverse. Out of the $53$ models plotted, $46$ points downwards and only $7$ upwards. The VDD model provides the key to understand this result.

\subsubsection{Hydrodynamical interpretation}
\label{sect:hydro}

Following \citet{haubois2012}, we computed the time evolution of VDD density profiles with the \texttt{SINGLEBE} code \citep{okazaki2007,okazaki2002}, and fed these to the LTE flux expression derived in Appendix~\ref{appendix:sed_evol} to compute the continuum fluxes at $12$, $25$ and $60\,\mathrm{{\mu}m}$ as a function of time. Then, the same procedure of Section~\ref{sect:sed_fit} was used to obtain from each synthetic SED a pair of $\log\rho_0$ and $n$ values from the pseudo-photosphere model.

\citet{haubois2012} explored the time-dependent VDD model predictions for three cases of interest: (i) a forming disc with a constant mass injection rate, (ii) a dissipating disc with no mass injection, and (iii) a disc subject to a periodic mass injection. For a forming disc with a constant mass injection rate, the density radial profile is initially very steep, but progressively approaches the steady-state value ($n=3.5$) during the disc build-up. Technically, a decretion disc never actually reaches the steady-state, since it takes an infinite time to do so \citep{okazaki2007}. However, as the disc build-up occurs from inside-out, the disc inner region approaches steady-state before the outer parts. Therefore, the disc observables that probe this inner region will appear similar to a steady-state disc (see Appendix~\ref{appendix:sed_evol}). For the case when fully developed discs have the mass injection rate suddenly turned-off, the material of the disc inner part is re-accreted due to the outward angular momentum transfer via the disc turbulent viscosity. The simultaneous infall in the inner disc and outflow in the outer disc gives rise to a stagnation radius in the disc, where the radial velocity is zero. This stagnation radius slowly propagates outward, and the density structure within this radius evolves by decreasing its density level while maintaining its radial profile \citep[][see also Appendix~\ref{appendix:sed_evol}]{haubois2012}.

%-------------------------------------------------
% comparison with other works
%-------------------------------------------------
%-------------------------------------------------

%-------------------------------------------------
% lum-mdot diagram
%-------------------------------------------------
%-------------------------------------------------

%-------------------------------------------------
% evolution akari - wise
%-------------------------------------------------
\begin{figure}%[t]
\begin{center}
\includegraphics[width=1.\linewidth]{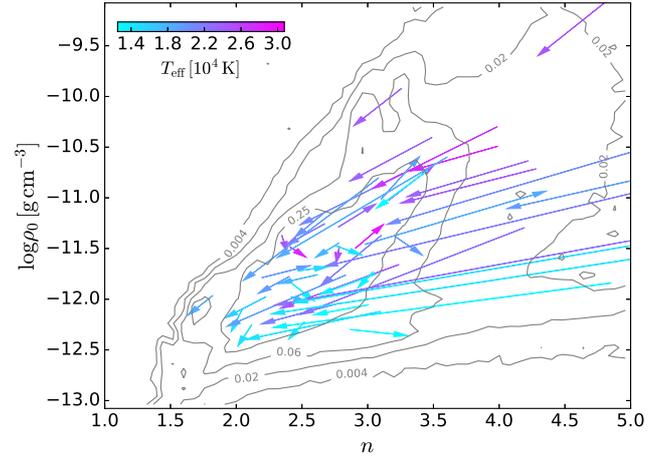}
\caption{Evolution of the disc parameters between different epochs. The arrow tails correspond to values derived from \textit{AKARI} observations, while the arrow heads correspond to determinations from \textit{AllWISE} data. The arrows were superimposed on the probability distribution of the fitted parameters (see Figure~\ref{fig:n_ln0_histo2d}), and their colors correspond to the effective temperature determined by \citet{fremat2005}. \label{fig:evol_akari_wise}}
\end{center}
\end{figure}
%-------------------------------------------------

%-------------------------------------------------
% evolutionary tracks
%-------------------------------------------------
\begin{figure}%[!t]
\begin{center}
\subfigure[Disc build-up and dissipation. In this case, the disc is fed for $30\,\mathrm{yr}$ at a constant rate (portion from the right to $n\simeq 3.4$ of the tracks), which is then turned-off for $60\,\mathrm{yr}$. The arrows indicate the evolution direction, and the time step was held fixed at 6~months. In the plot, the steps where $Z_{12\,\mathrm{\mu m}}-1<0.05$ were omitted.]{%
\label{fig:track_buildup_dissip}
\includegraphics[width=1.\linewidth]{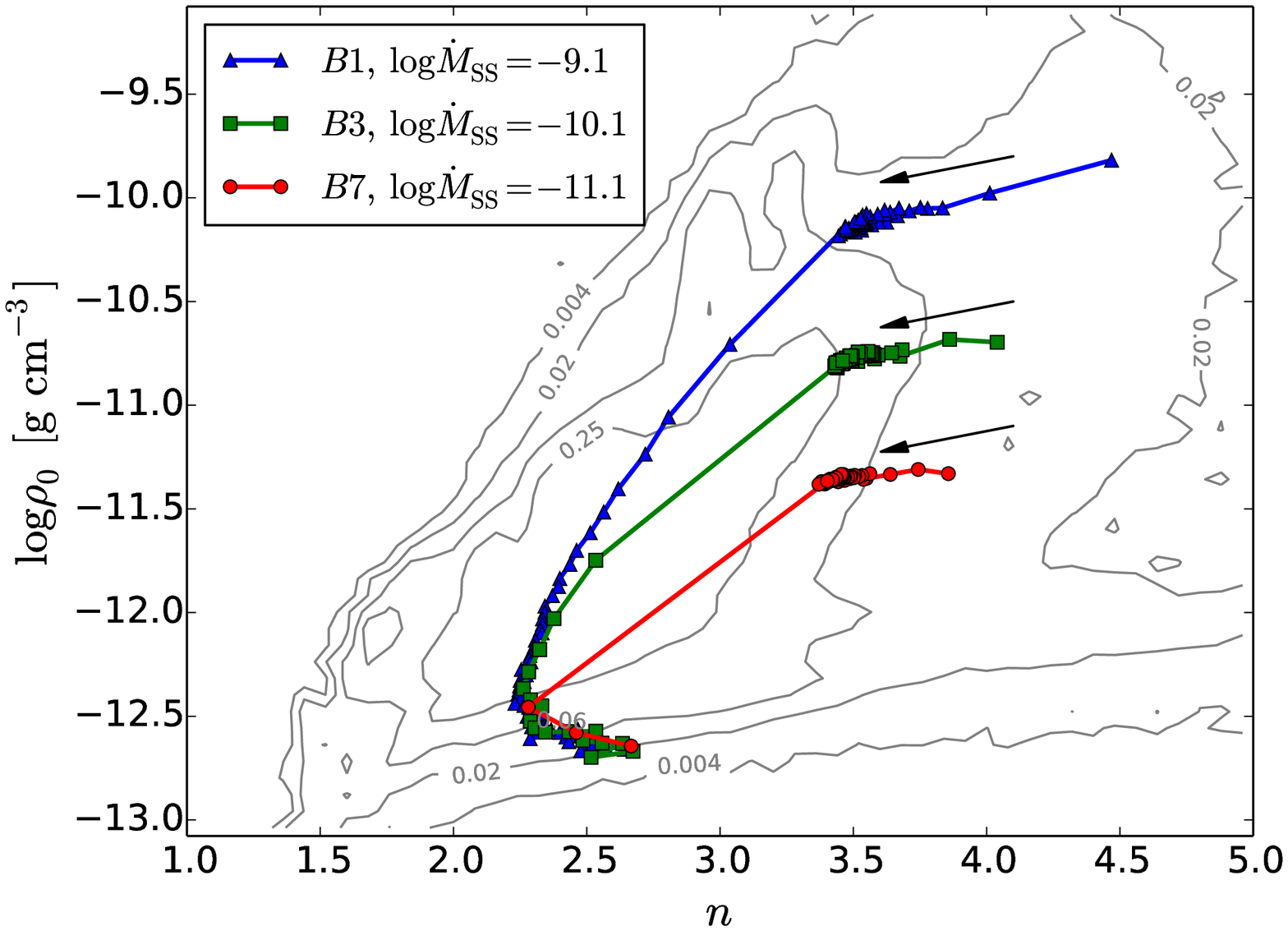}
}\\
\subfigure[Disc subjected to a periodic mass injection. The adopted period was $5\,\mathrm{yr}$, with a constant mass injection rate for half of each cycle and none the other half. The time step was held fixed at 6~months, and the $11^{\mathrm{th}}$ cycle is shown (i.e., the initial time corresponds to 50~yr). The arrows indicate both track initial positions and direction of evolution.]{%
\label{fig:track_periodic}
\includegraphics[width=1.\linewidth]{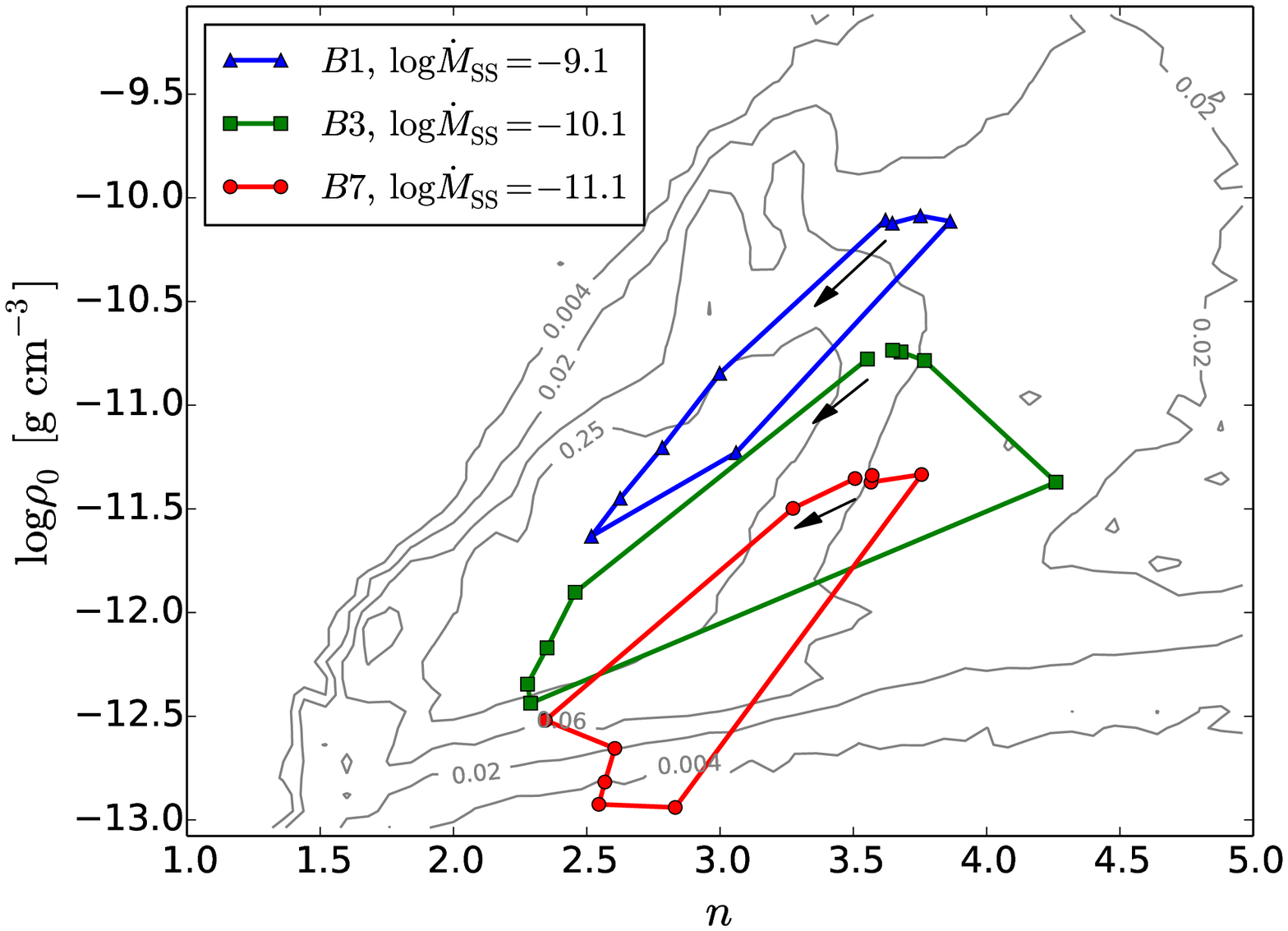}
}\\
\end{center}
\caption{Disc evolution across the $n-\log\rho_0$ diagram. The evolutionary tracks are superimposed on the probability distribution of the fitted parameters (see Figure~\ref{fig:n_ln0_histo2d}). The disc models have viscosity parameter $\alpha=1$ and pole-on orientation. The adopted mass decretion rates and spectral types are indicated.}
\label{fig:tracks}
\end{figure}
%-------------------------------------------------

Figure~\ref{fig:tracks} shows the computed evolutionary tracks across the $n-\log\rho_0$ diagram (see Appendix~\ref{appendix:sed_evol}). In Figure~\ref{fig:track_buildup_dissip}, during the disc build-up, the disc parameters reach the steady-state strip ($n\sim 3.5$) in less than one year. Subsequently, the solution remains close to $n=3.5$ as long as mass is provided to the disc. When the mass injection is turned-off, the inner disc quickly dissipates and the evolutionary tracks move toward the left-bottom position of the diagram. This direction coincides with the results seen in Figure~\ref{fig:evol_akari_wise}, suggesting that most of the observed discs are in a dissipating state. Furthermore, disc dissipation takes a longer time than that required for its build-up (when fed by a constant mass injection rate), which makes dissipating discs more likely to be observed than forming ones. This explains why more arrows in Figure~\ref{fig:evol_akari_wise} point toward the lower left. Both build-up and dissipation time scales are expected to be comparable only in the case of small discs, since the dissipation time scale increases with disc size \citep[e.g.,][]{oktariani2016}.

Another situation of interest is shown in Figure~\ref{fig:track_periodic}, where the disc is subjected to a periodic mass injection rate. The disc recovery is very fast, so the disc parameters rapidly reach the upper right portion of the loop. Again, the track asymptotically approaches the $n=3.5$ steady-state value while material is fed to the disc. The subsequent part of the disc dissipation then produces the slower excursion along the lower left part of the diagram. A loop-like behaviour can be also found for many other pairs of measurements probing the disc, such as photometry in different bands \citep{haubois2012}, polarimetry and Balmer discontinuity \citep{haubois2014}, and even interferometry \citep{faes_inprep}.

Finally, it is useful to recall that the association of $n=3.5$ to the disc steady-state is based on some model simplifications, such as disc isothermality and constant $\alpha$ (as a function of both position and time), and also assumes an isolated system \citep[e.g.,][]{bjorkman2005}. The change in the steady-state density exponent could be due to: (i) either an $\alpha$ and/or a $T_{\mathrm{d}}$ radial dependence \citep[e.g., ][Equation~24]{carciofi2008}; and (ii) the accumulation effect caused by a binary companion \citep{panoglou2016,klement2015,okazaki2002}, which can reduce $n$ to $\sim$$3$. For example, \citet{klement2015} found evidence for disc truncation in $\beta$~CMi with $n=3$. For these reasons, we expect steady-state discs to have $n$ in the range $3\lesssim n\lesssim 3.5$.

\subsection{The viscous disc life-cycle}
\label{sect:lifecycle}

\begin{figure}%[t]
\begin{center}
\includegraphics[width=1.\linewidth]{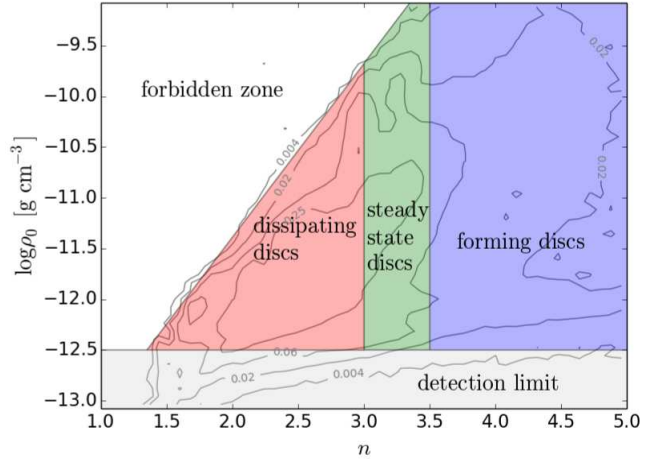}
\caption{Typical regions for distinct disc dynamical states in the $n-\log\rho_0$ diagram, according to VDD hydrodynamical simulations. The defined regions were superimposed on the combined {\ttfamily emcee} posterior distributions for our sample (see Figure~\ref{fig:n_ln0_histo2d}). \label{fig:track_regions}}
\end{center}
\end{figure}

From the representative examples shown in Figure~\ref{fig:tracks}, the time-dependent behaviour of the disc can be summarized in terms of regions in the $n-\log\rho_0$ diagram. Figure~\ref{fig:track_regions} shows the definition of such regions, divided into: forming discs ($n\gtrsim 3.5$); steady-state discs ($3\lesssim n\lesssim 3.5$) and dissipating discs ($n\lesssim 3$). Since no stars are observed in the upper left region of the diagram, it is called the ``forbidden zone''. Reaching this region would require either a (not observed) much higher $\dot{M}_{\mathrm{SS}}$ ($\gtrsim 10^{-8}\,\mathrm{M_{\odot}\,yr^{-1}}$), or the presence of a closer/more massive binary companion. Finally, the detection limit was defined as the line below which the flux excess emerging from the disc becomes negligible, as shown in Figure~\ref{fig:deltmag}. This may be not the case for other disc observables (e.g., emission lines), which eventually can probe smaller densities.

From a total of $169$ fitted models, $93$ ($55\%$) populate the dissipating region, $40$ ($24\%$) are in the formation region. In this last group, $17$ ($10\%$ of the total sample) have $n>5$, which may also indicate a disc-less state. These numbers suggest that, on average for our sample, the disc dissipation takes $\sim$$2$ times longer than the disc build-up. \citet{rivinius1998a} found a similar behaviour for the line emission variability of $\mu$~Cen, where the observed outbursts are followed by extended relaxation phases. Similar results were found by \citet{haubois2012} and \citet{huat2009}. In both cases, the dissipation time scales appear to be longer than the outburst episodes. Finally, only $36$ models ($21\%$) are found in the steady-state strip. According to the theoretical evolutionary tracks in Figure~\ref{fig:tracks}, the discs must be fed for long enough to remain in the steady-state region. The smaller number of discs observed in the steady dynamical state suggests that the mass injection episodes are probably shorter than the duration of the dissipation phase, and therefore less likely to be observed.

\section{Conclusions}
\label{sect:conclusions}

%-------------------------------------------------
% track regions
%-------------------------------------------------
%-------------------------------------------------

We have systematically applied the VDD model to measure the disc properties of a large sample of Be stars using their IR SEDs. The MCMC method provided a reliable determination of the disc parameters of this sample at different epochs, producing a total of 169 models. The combination of the posterior probability distributions computed with \texttt{emcee} MCMC implementation showed that the density exponent mainly lies between $1.5$ and $3.5$, while the disc base density of $\log\rho_0\simeq -12$ is the most probable to be found among Be discs. The disc base densities typically range from this lower limit up to $\log\rho_0\simeq-10$, and we observe a positive correlation between $n$ and $\log\rho_0$. Such results are in agreement with those found by \citet{waters1987}.

Denser discs are more likely to occur around earlier-type Be stars, which either means that more massive objects can provide higher mass injection rates to the disc, or hotter objects have smaller $\alpha$. This question requires further investigation. No clear correlation was found between $n$ and $T_{\mathrm{eff}}$.

The disc mass decretion rates were estimated for our sample under the approximation of a steady-state disc. The values for $\dot{M}_{\mathrm{SS}}/\alpha$ range from $10^{-12}$ to $10^{-8}\,\mathrm{M_{\odot}\, yr^{-1}}$, and they increase with the stellar luminosity. These results are up to three orders of magnitude smaller than those found by \citet{waters1987}, since the VDD outflow velocities are much smaller than their outflowing wind model. Additionally, our results do not show a mass-loss transition at $\log(L_{\star}/\mathrm{L_{\odot}})=3$, as suggested by those authors. Finally, our results are compatible with the mass decretion rates found by \citet{granada2013}, which were based on stellar evolution arguments.

The dynamical scenario predicted by the time-dependent VDD model provides a satisfactory interpretation key for the results from the SED fitting. Evolutionary tracks on the $n-\log\rho_0$ diagram suggest that most of the observed cases correspond to dissipating discs, since this evolutionary stage has a much longer time scale than the disc formation stage.
The hydrodynamical interpretation of our results leads to the classification of distinct regions of the $n-\log\rho_0$ plane, associated with different evolutionary stages. The $n\lesssim 3$ region is associated with flatter SED slopes and dissipating discs; the strip between $n\simeq 3$ and $n\simeq 3.5$ corresponds to the steady-state zone; finally, forming discs occupy the region where $n\gtrsim 3.5$. The more extended region for the steady-state rather than the canonical $n=3.5$ may be due to disc non-isothermality and/or non-isoviscosity. The smaller number of solutions around the steady-state region suggests the discs spend less time being actively fed than passively dissipating, which means that mass injection episodes must be shorter when compared to the dissipation time. Studies with a denser time coverage are required in order to impose better constraints on such time scales.

% ACKNOWLEDGEMENTS________________________________

\section*{Acknowledgments}

We thank the anonymous referee for the useful comments. This work made use of the computing facilities of the Laboratory of Astroinformatics (IAG/USP, NAT/Unicsul), whose purchase was made possible by the Brazilian agency FAPESP (grant 2009/54006-4) and the INCT-A. 
%; the BeSS database, operated at LESIA, Observatoire de Meudon, France: \url{http://basebe.obspm.fr}; the VizieR catalogue access tool, CDS, Strasbourg, France; VOSA, developed under the Spanish Virtual Observatory project supported from the Spanish MICINN through grant AyA2011-24052.
R.~G.~V. acknowledges the support from FAPESP (grant 2012/20364-4), A. C. C acknowledges support from CNPq (grant 307594/2015-7) and FAPESP (grant 2015/17967-7), J.~E.~B. acknowledges support from the NSF (grant AST-1412135).

% BIBLIOGRAPHY___________________________________

% APPENDIX________________________________________

\appendix

\section{Selected sample}
\label{appendix:sample}

Table~\ref{tab:sample} lists the selected Be stars from \citeauthor{fremat2005}'s sample, their respective spectral classification and color-corrected fluxes (see Section~\ref{sect:sample}).

\section{Computing the stellar parameters}
\label{appendix:stellar_par}

\citet{fremat2005} derived the fundamental parameters of 130 $B$-type stars, based on the fit of their optical spectra. These authors made a distinction between \textit{apparent} parameters, where the rotation effects are neglected, and \textit{parent non-rotating counterpart} (pnrc) parameters, which span a family of models of different rotation velocities.
In this work, we adopted the pnrc parameters to compute our SED models. However, they have first to be converted to the fundamental parameters of a specific rotation rate. The pnrc parameters are $T_{\mathrm{eff,\, 0}}$, $\log g_{\mathrm{0}}$, the true $v\sin i$, the critical velocity $V_{\mathrm{c}}$ and the inclination $i$, while the rotating stellar parameters necessary for the pseudo-photosphere model are the rotation rate $\omega = \Omega/\Omega_{\mathrm{crit}}$ (which is equivalent to $W$, see \citealt{rivinius2013}), $M_{\star}$, $R_{\mathrm{pole}}$ and $L_{\star}$. In this appendix, we derive the expressions needed to convert the pnrc parameters into the rotating model parameters of interest. By evaluating Equation~(2) from \citet{fremat2005} at $\omega = 1$, we have:
\begin{equation}\label{eq:rpole}
R_{\mathrm{pole}}(\omega = 1) = R_{\mathrm{0}} [1 - 0.0152\, P(M_{\star})],
\end{equation}
where $R_{\mathrm{0}}$ is the radius of the non-rotating star, and
\begin{align}
P(M_{\star}) &= 5.66 + \frac{9.43}{(M_{\star}/\mathrm{M_{\odot}})^2} & (M_{\star}/\mathrm{M_{\odot}} \gtrsim 2).
\end{align}
The rotational critical velocity is given by \citep[e.g.,][]{rivinius2013}
\begin{equation}
V_{\mathrm{c}}^2 = \frac{2}{3}\frac{G M_{\star}}{\left. R_{\mathrm{pole}}\right|_{\omega=1}} = \frac{2}{3}\frac{G M_{\star}}{R_{\mathrm{0}} [1 - 0.0152\, P(M_{\star})]},
\end{equation}
and the stellar mass can be written as
\begin{equation}
M_{\star} = \frac{g_{\mathrm{0}} R_{\mathrm{0}}^2}{G}.
\end{equation}
By using these definitions in Equation~(\ref{eq:rpole}), we find an equation for $R_{\mathrm{0}}$:
\begin{equation}
R_{\mathrm{0}} = \frac{3}{2}\frac{V_{\mathrm{c}}^2}{g_{\mathrm{0}}} \left[1 - 0.0152\, P\left(\frac{g_{\mathrm{0}} R_{\mathrm{0}}^2}{G}\right)\right],
\end{equation}
which can be numerically solved. Once $R_{\mathrm{0}}$ is known, we can compute $R_{\mathrm{pole}}$ from Equation~(\ref{eq:rpole}), and use Equation~(1) from \citet{fremat2005} to derive an expression for the equatorial radius:
\begin{equation}
\frac{R_{\mathrm{eq}}/R_{\mathrm{0}}}{1-P(M_{\star})\,\tilde{\tau}} = 1 + \frac{\omega^2}{2}\left[ \frac{R_{\mathrm{eq}} / (1.5\,R_{\mathrm{0}})}{1-0.0152P(M_{\star})} \right]^3,
\end{equation}
where
\begin{equation}
\omega = \cos\left\{3 \left[\cos^{-1}\left(\frac{v_{\mathrm{rot}}}{2 V_{\mathrm{c}}}\right) - \pi\right]\right\}
\end{equation}
\citep[e.g., ][]{rivinius2013},
\begin{equation}
\tilde{\tau} \simeq (0.0072 + 0.008 \eta^{1/2})\,\eta^{1/2},
\end{equation}
and
\begin{equation}
\eta = \omega^2\,\left\{\frac{R_{\mathrm{eq}}}{1.5\,R_0\,[1 - 0.0152\, P(M_{\star})]}\right\}^3.
\end{equation}
Finally, the luminosity can be computed by:
\begin{equation}
L_{\star} = L_0\,\left[\tilde{a} + (1-\tilde{a})\,\mathrm{e}^{-\tilde{b}\tilde{\tau}}\right],
\end{equation}
where
\begin{equation}
L_0 = 4\pi R_0^2\, \sigma_{\mathrm{B}}  T_{\mathrm{eff,\,0}}^4,
\end{equation}
\begin{equation}
\tilde{a} = 0.675 + 0.046 \left(\frac{M_{\star}}{\mathrm{M_{\odot}}}\right)^{1/2},
\end{equation}
\begin{equation}
\tilde{b} = 52.71 + 20.63 \left(\frac{M_{\star}}{\mathrm{M_{\odot}}}\right)^{1/2}.
\end{equation}

\section{MCMC results}
\label{appendix:results}

Table~\ref{tab:results} presents the stellar and disc parameters derived with the \texttt{emcee} code. The parameter values correspond to the median of the sampled distributions, and the derived uncertainties correspond to a 1-$\sigma$ confidence interval (see Section~\ref{sect:results}).

\section{Effective radii and disc truncation effects}
\label{appendix:ref_trunc}

\begin{figure}%[t]
\begin{center}
\includegraphics[width=1.\linewidth]{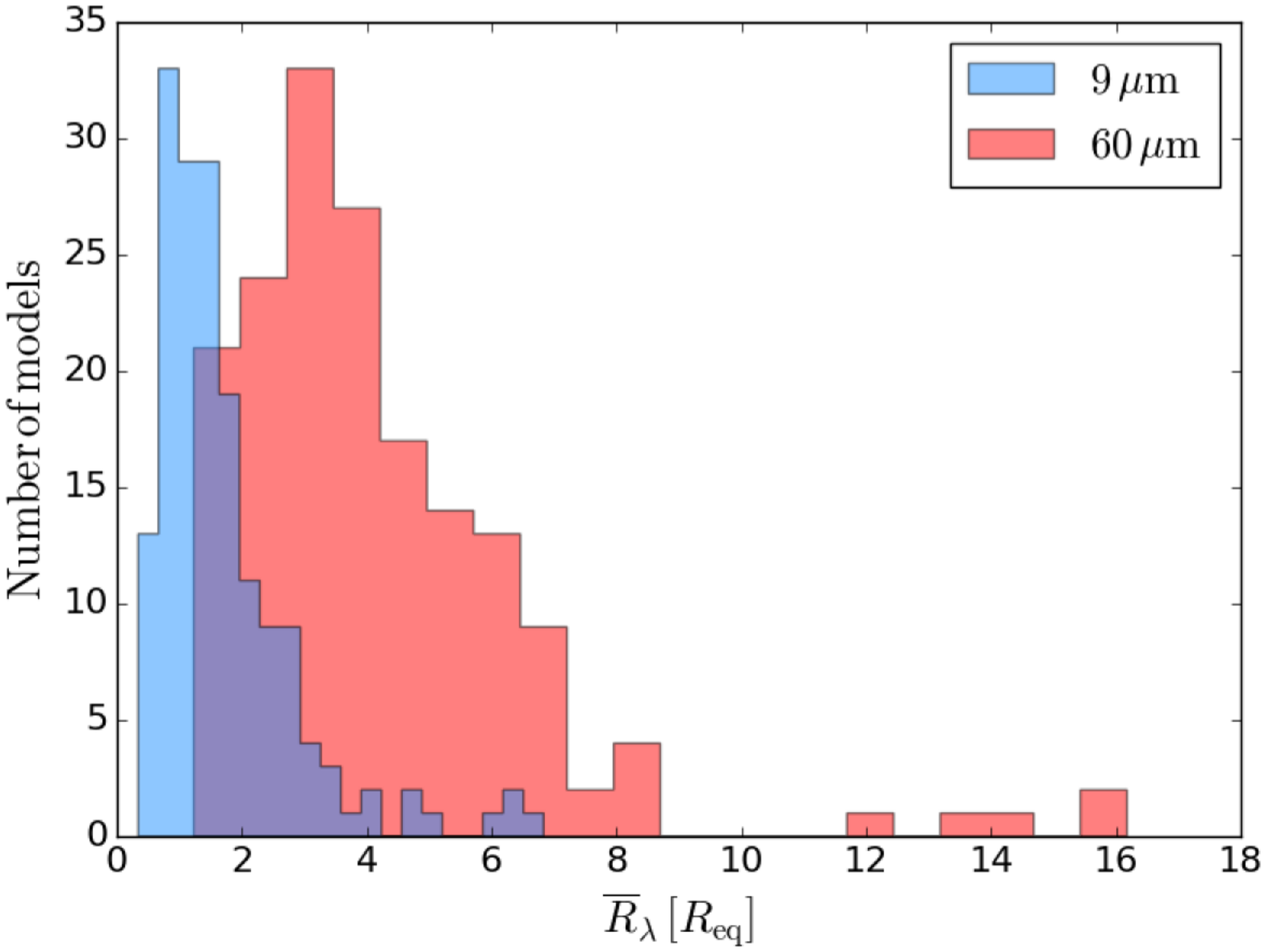}
\caption{Distribution of derived $\overline{R}$ values, computed at the shortest and longest wavelengths adopted in this work. \label{fig:ref_hist}}
\end{center}
\end{figure}

The disc effective radius is a function of wavelength, and thus has a particular value at each IR bandpass adopted for this work. The interested reader can easily compute it using the following expression:
\begin{align}
\frac{\overline{R}}{R_{\mathrm{eq}}} &= \left[ 1.33\times 10^{-3}\,\left(\frac{T_{\mathrm{eff}}}{10^4\,\mathrm{K}}\right)^{-1}\,\left(\frac{R_{\mathrm{eq}}}{R_{\odot}}\right)^{3/2}\,\left(\frac{M_{\star}}{M_{\odot}} \right)^{-1/2} \right. \\
& \left. \quad\times \left(\frac{\rho_0}{10^{-12}\,\mathrm{g\,cm^{-2}}} \right)^2\, \lambda_{\mathrm{{\mu}m}}^2\, \left(g_{\mathrm{ff}} + g_{\mathrm{bf}} \right)\right]^{1/(2n-\beta)},
\end{align}
where the same assumptions described in Section~\ref{sect:sed_fit} were adopted. Figure~\ref{fig:ref_hist} shows the $\overline{R}$ distribution for the shortest and the longest wavelengths used in this work. Although the $\overline{R}/R_{\mathrm{eq}}<1$ cases have no geometrical interpretation, they still remain useful for providing information about the disc vertical optical depth scale \citep{vieira2015}. We see that $\overline{R}/R_{\mathrm{eq}}<10$ for practically all the sample (except for HD~$41335$ and HD~$110432$ at $60\,\mathrm{{\mu}m}$).

Disc truncation caused by a binary companion may be of importance for the SED if it disrupts the pseudo-photosphere at the wavelength of interest. The main result of disc truncation is a discontinuity in the SED first derivative at $\overline{\lambda}$ such that $\overline{R}(\overline{\lambda})=R_{\mathrm{t}}$, where $R_{\mathrm{t}}$ is the truncation radius (\citealt{vieira2015}; see also Fig.~27 in \citealt{panoglou2016}). At $\lambda>\overline{\lambda}$, the SED slope becomes equal to the photospheric one.

It follows that the results presented in this paper can potentially be affected by the presence of unknown binaries if the truncation radius is $R_{\mathrm{t}}\lesssim 10\,R_{\mathrm{eq}}$.

However, according to the work of \citet{panoglou2016} such a small truncation radius would be associated to short-period binaries ($P < 20$~days). The shortest known orbital periods in our sample are the Be$+$sdO systems $o$~Puppis \citep{koubsky2012} and $59$~Cyg \citep{rivinius2000}, both having $P\simeq 30$~days. Although the existence of an undetected companion so close to the Be cannot be discarded, it is probably unlikely to find such a dramatic case in the sample. The lack of evidence of disc truncation at relatively small radii is another hint that close binaries are not common among Be stars, and therefore binarity is probably not relevant for the ejection and formation of discs.

\section{SED evolution}
\label{appendix:sed_evol}

\begin{figure*}
\begin{center}
\subfigure[Disc build-up, with a constant mass injection rate of $3.9\times 10^{-8}\,\mathrm{M_{\odot}\,yr^{-1}}$.]{%
\label{fig:buildup}
\includegraphics[width=.85\linewidth]{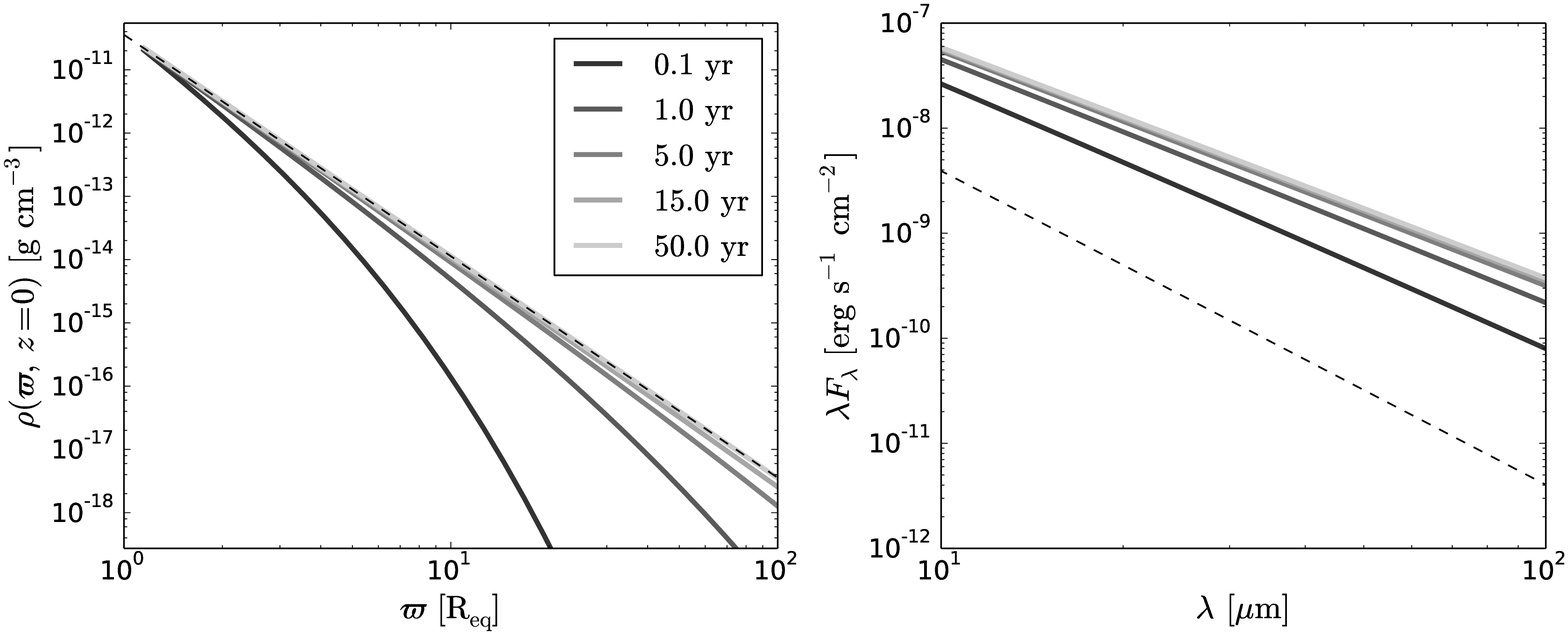}
}\\
\subfigure[Disc dissipation, departing from steady-state with base density of $2.7\times 10^{-11}\,\mathrm{g\,cm^{-3}}$.]{%
\label{fig:dissipation}
\includegraphics[width=.85\linewidth]{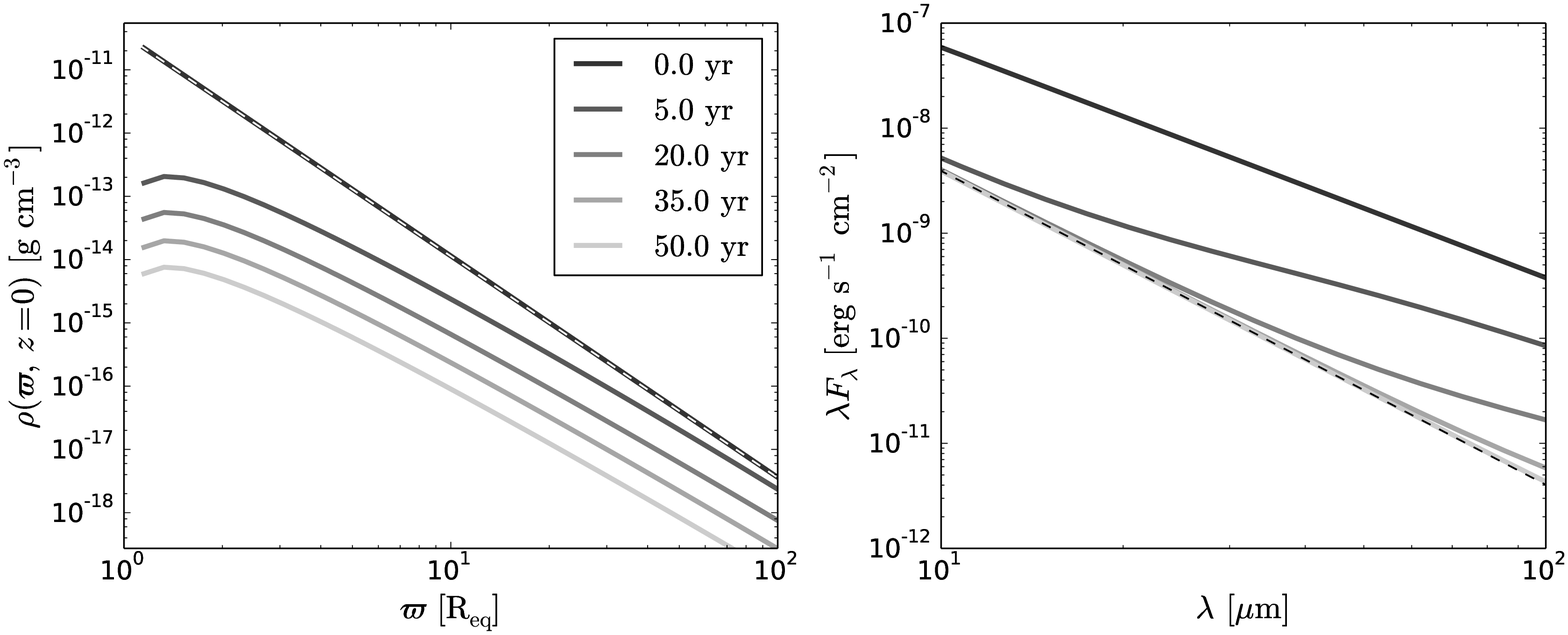}
}\\
\subfigure[Periodic mass injection with a mass injection rate of $7.7\times 10^{-9}\,\mathrm{M_{\odot}\,yr}$, a period of $5\,\mathrm{yr}$ and duty-cycle of $50\,\%$. The mass injection rate is shown in the subplot, where the selected phases ($\phi$) are indicated. The chosen cycle corresponds to the $10^{\mathrm{th}}$ period, when the oscillatory disc properties are already stable.]{%
\label{fig:periodic}
\includegraphics[width=.85\linewidth]{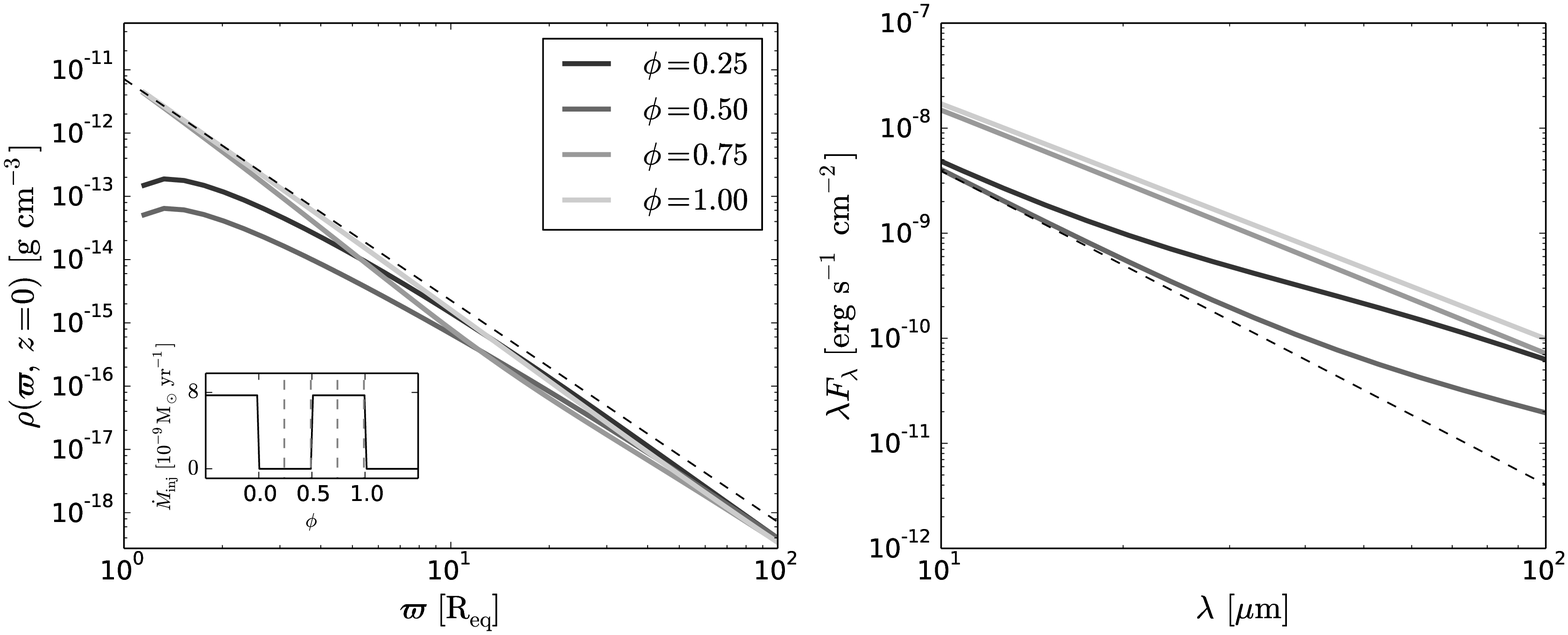}
}\\
\end{center}
\caption{Density profile at disc mid-plane and SED evolution for different mass injection scenarios. The dashed lines in the left panels correspond to the steady-state profile, while the dashed lines in the right panels indicate the stellar photospheric flux. For all cases, the central star has a \textit{B2} spectral type (see Table~\ref{tab:stellar_par}), $d=50\,\textrm{pc}$ and a pole-on orientation.}
\label{fig:sig_sed}
\end{figure*}

The code \texttt{SINGLEBE} \citep{okazaki2007, okazaki2002} computes the one-dimensional time evolution of the radial density profile of a decretion disc. Mass is injected into orbit above the base of the disc at $1.04\,R_{\mathrm{eq}}$, and the disc is assumed to be azimuthally symmetric. Since this mass injection is the source of the angular momentum carried away by the decretion disc, most of the injected mass falls back onto the star. Typically, the decretion rate of the disc, $\dot{M}_{\mathrm{SS}}$, is two orders of magnitude smaller than the injection rate. Given the disc density profile, the free-free and bound-free LTE opacities can be written as \citep{brussaard1962}
\begin{align}\label{kappa}
\kappa_{\lambda} &= 3.692\times 10^8\, \left[1-\exp\left(-hc/\lambda k T_{\textrm{d}}\right)\right]\overline{z^2} T_{\textrm{d}}^{-1/2} \nonumber \\
                 & \quad\times\left(\lambda/c\right)^{3} \gamma \left(\rho/\mu\,m_{\textrm{H}}\right)^2 \left[g_{\mathrm{ff}}(\lambda,T_{\textrm{d}}) + g_{\mathrm{bf}}(\lambda,T_{\textrm{d}})\right],
\end{align}
where we again use the isothermal approximation (see Section~\ref{sect:pphot_model}). For simplicity, we restrict our discussion of the SED evolution to the pole-on case. The vertical optical depth is given by
\begin{equation}
\tau_{\mathrm{z}} = \int_{-\infty}^{+\infty}\kappa_{\lambda}\,dz = \tau_0 \, \frac{H(\varpi)}{H_0} \, \left[\frac{\rho(\varpi, z=0)}{\rho_0}\right]^2,
\end{equation}
where $H(\varpi)/H_0=\varpi^{3/2}$ is the isothermal scale height, and
\begin{equation}\label{tau0}
\tau_0=\sqrt{\pi}\,H_0\,\kappa_{\lambda}(\varpi=R_{\mathrm{eq}},z=0).
\end{equation}
The specific intensity can be expressed as
\begin{equation}\label{bright_general}
I_{\lambda}(\varpi)=
\begin{cases}
S^{\star}_{\lambda}(\langle T_{\textrm{eff}}\rangle) & \textrm{($\varpi\leq R_{\mathrm{eq}}$)}\\
B_{\lambda}(T_{\textrm{d}})\,[1-\exp(-\tau_{\mathrm{z}})] & \textrm{($R_{\mathrm{eq}}<\varpi\leq R_{\mathrm{d}}$),}\\
\end{cases}
\end{equation}
and consequently the total flux can be expressed by
\begin{equation}
F_{\lambda}=\frac{1}{d^2}\int_{0}^{R_{\mathrm{d}}} I_{\lambda}(\varpi)\,2\pi \varpi d\varpi.
\end{equation}
Figure~\ref{fig:sig_sed} shows the evolution of both the disc radial density profile and the IR SED for some dynamical scenarios of interest. The SED excess responds very quickly to the disc formation, affecting the entire IR wavelength range within a few weeks (upper panels). In contrast, the disc dissipation slowly modifies the SED with the excess decreasing first at shorter wavelengths (middle panels). Finally, the bottom panels show the case of periodic mass injection.

%-------------------------------------------------
% Ref histo
%-------------------------------------------------
%\afterpage{
%}
%-------------------------------------------------

%\pagebreak
\newpage

%-------------------------------------------------
% Table sample
%-------------------------------------------------
\afterpage{
\nonstopmode
\setlongtables
\onecolumn
\begin{landscape}
\begin{longtable}{lllccccccc}
\kill
\caption{Selected sample and respective color-corrected fluxes.} \label{tab:sample}
\\
\hline
Name & HD & Specral type$^a$ & \multicolumn{3}{c}{IRAS} & \multicolumn{2}{c}{AKARI} & \multicolumn{2}{c}{WISE}\\
 & & & $12\,\mathrm{{\mu}m}$ & $25\,\mathrm{{\mu}m}$ & $60\,\mathrm{{\mu}m}$ & $9\,\mathrm{{\mu}m}$ & $18\,\mathrm{{\mu}m}$ & $12\,\mathrm{{\mu}m}$ & $22\,\mathrm{{\mu}m}$\\
 & & & $\mathrm{[Jy]}$ & $\mathrm{[Jy]}$ & $\mathrm{[Jy]}$ & $\mathrm{[Jy]}$ & $\mathrm{[Jy]}$ & $\mathrm{[Jy]}$ &$\mathrm{[Jy]}$\\ \hline
$\gamma$~Cas & 5394 & B0.5IVpe &$14.9\pm 1.1$ & $6.7\pm 0.7$ & $2.2\pm 0.5$ & $28.7\pm 0.1$ & $13.02\pm 0.04$ &  -- & -- \\ 
$\phi$~And & 6811 & B5IIIe &$0.8\pm 0.1$ & $0.4\pm 0.1$ & -- &  -- & -- &  -- & -- \\ 
HD~11606 & 11606 & B2Vne &$0.5\pm 0.1$ & $0.25\pm 0.04$ & -- & $0.47\pm 0.05$ & $0.25\pm 0.02$ & $0.321\pm 0.004$ & $0.190\pm 0.004$ \\
HD~18552 & 18552 & B7IVe & -- & -- & -- & $0.32\pm 0.01$ & $0.17\pm 0.01$ & $0.215\pm 0.003$ & $0.123\pm 0.003$ \\
HD~20336 & 20336 & B2.5Vn(e) & -- & -- & -- & $1.6\pm 0.1$ & $0.75\pm 0.01$ & $1.28\pm 0.01$ & $0.75\pm 0.01$ \\
HD~22780 & 22780 & B7Vn & -- & -- & -- &  -- & -- & $0.404\pm 0.005$ & $0.189\pm 0.004$ \\
$13$~Tau & 23016 & B9Vne & -- & -- & -- &  -- & -- & $0.150\pm 0.002$ & $0.048\pm 0.002$ \\
$23$~Tau & 23480 & B6IVe & -- & -- & -- & $1.46\pm 0.03$ & $1.3\pm 0.1$ & $0.68\pm 0.01$ & $0.53\pm 0.02$ \\
HD~23552 & 23552 & B8V & -- & -- & -- & $0.38\pm 0.02$ & $0.23\pm 0.02$ & $0.225\pm 0.003$ & $0.127\pm 0.003$ \\
$\eta$~Tau & 23630 & B7III &$3.4\pm 0.2$ & $1.7\pm 0.2$ & -- & $4.3\pm 0.1$ & $2.0\pm 0.1$ & $3.3\pm 0.1$ & $1.52\pm 0.03$ \\
V*~X~Per & 24534 & O9.5III & -- & -- & -- & $1.9\pm 0.1$ & $0.76\pm 0.01$ & $1.41\pm 0.01$ & $0.70\pm 0.01$ \\
$48$~Per & 25940 & B3Ve &$3.1\pm 0.2$ & $1.6\pm 0.2$ & $0.7\pm 0.1$ & $4.01\pm 0.03$ & $2.1\pm 0.1$ & $3.08\pm 0.02$ & $1.74\pm 0.03$ \\
DU~Eri & 28497 & B2(V)ne &$0.8\pm 0.1$ & $0.4\pm 0.1$ & -- & $0.86\pm 0.03$ & $0.38\pm 0.04$ & $0.74\pm 0.01$ & $0.34\pm 0.01$ \\
$11$~Cam & 32343 & B3Ve &$1.6\pm 0.2$ & $1.3\pm 0.1$ & $2.2\pm 0.2$ & $1.15\pm 0.03$ & $0.8\pm 0.1$ & $0.87\pm 0.01$ & $0.68\pm 0.01$ \\
$\lambda$~Eri & 33328 & B2IVne & -- & -- & -- &  -- & -- & $0.72\pm 0.01$ & $0.30\pm 0.01$ \\
$\psi^1$~Ori & 35439 & B1Vn &$1.2\pm 0.1$ & $0.5\pm 0.1$ & -- & $1.6\pm 0.1$ & $0.68\pm 0.04$ & $1.45\pm 0.02$ & $0.68\pm 0.01$ \\
$120$~Tau & 36576 & B2IV-Ve &$1.1\pm 0.1$ & $0.9\pm 0.3$ & $1.3\pm 0.2$ & $1.75\pm 0.03$ & $1.12\pm 0.03$ & $1.25\pm 0.01$ & $0.82\pm 0.02$ \\
HD~37657 & 37657 & B3Vne & -- & -- & -- & $0.37\pm 0.02$ & $0.19\pm 0.01$ & $0.118\pm 0.002$ & $0.090\pm 0.002$ \\
$\alpha$~Col & 37795 & B9Ve &$4.7\pm 0.2$ & $2.3\pm 0.2$ & $0.9\pm 0.1$ & $5.16\pm 0.02$ & $2.58\pm 0.02$ & $3.92\pm 0.04$ & $2.18\pm 0.03$ \\
HD~37967 & 37967 & B2.5Ve & -- & -- & -- & $0.87\pm 0.01$ & $0.469\pm 0.001$ & $0.61\pm 0.01$ & $0.38\pm 0.01$ \\
V*~V1165~Tau & 38010 & B1Vpe &$0.6\pm 0.1$ & $0.5\pm 0.1$ & -- & $0.70\pm 0.04$ & $0.46\pm 0.05$ & $0.46\pm 0.01$ & $0.28\pm 0.01$ \\
HD~40978 & 40978 & B3Ve & -- & -- & -- &  -- & -- & $0.164\pm 0.002$ & $0.089\pm 0.002$ \\
HD~41335 & 41335 & B3/5Vnne &$1.8\pm 0.2$ & $0.9\pm 0.1$ & -- &  -- & -- & $1.47\pm 0.02$ & $0.77\pm 0.01$ \\
HD~44458 & 44458 & B1.5IVe &$1.4\pm 0.2$ & $0.7\pm 0.1$ & -- & $1.40\pm 0.05$ & $0.7\pm 0.1$ & $0.83\pm 0.01$ & $0.50\pm 0.01$ \\
HD~45995 & 45995 & B1.5Vne & -- & -- & -- & $0.75\pm 0.01$ & $0.3\pm 0.1$ & $0.390\pm 0.005$ & $0.237\pm 0.005$ \\
HD~47054 & 47054 & B8IVe & -- & -- & -- & $0.45\pm 0.03$ & $0.3\pm 0.1$ & $0.325\pm 0.004$ & $0.181\pm 0.004$ \\
$\kappa$~CMa & 50013 & B1.5Ve &$5.2\pm 0.4$ & $2.4\pm 0.2$ & $0.8\pm 0.1$ & $5.4\pm 0.2$ & $2.5\pm 0.1$ & $5.33\pm 0.02$ & $2.48\pm 0.03$ \\
$\omega$~CMa & 56139 & B2IV-Ve &$1.5\pm 0.2$ & $0.7\pm 0.1$ & $0.4\pm 0.1$ & $2.7\pm 0.3$ & $1.7\pm 0.2$ & $1.88\pm 0.01$ & $0.88\pm 0.01$ \\
HD~58050 & 58050 & B2Ve & -- & -- & -- &  -- & -- & $0.046\pm 0.001$ & $0.017\pm 0.001$ \\
HD~58343 & 58343 & B2Vne &$1.0\pm 0.1$ & $0.6\pm 0.1$ & -- &  -- & -- & $0.41\pm 0.01$ & $0.30\pm 0.01$ \\
$\beta$~CMi & 58715 & B8Ve &$3.2\pm 0.2$ & $1.5\pm 0.2$ & $0.6\pm 0.1$ & $4.44\pm 0.01$ & $2.1\pm 0.1$ & $3.16\pm 0.03$ & $1.60\pm 0.02$ \\
HD~58978 & 58978 & B0.5IVe &$0.7\pm 0.1$ & $0.24\pm 0.04$ & -- &  -- & -- & $0.225\pm 0.003$ & $0.108\pm 0.003$ \\
HR~2911 & 60606 & B2Vne &$1.2\pm 0.2$ & $0.7\pm 0.1$ & $0.7\pm 0.1$ & $1.05\pm 0.02$ & $0.71\pm 0.04$ & $0.74\pm 0.01$ & $0.45\pm 0.01$ \\
HD~60848 & 60848 & O8:V: & -- & -- & -- &  -- & -- & $0.100\pm 0.001$ & $0.049\pm 0.002$ \\
$o$~Pup & 63462 & B1IVe &$2.0\pm 0.2$ & $0.7\pm 0.1$ & -- &  -- & -- & $1.81\pm 0.02$ & $0.72\pm 0.01$ \\
HD~65875 & 65875 & B2.5Ve & -- & -- & -- & $0.51\pm 0.02$ & $0.28\pm 0.01$ & $0.38\pm 0.01$ & $0.229\pm 0.005$ \\
HR~3237 & 68980 & B2ne &$1.7\pm 0.1$ & $1.0\pm 0.1$ & $1.0\pm 0.1$ & $1.91\pm 0.05$ & $0.99\pm 0.02$ & $1.39\pm 0.02$ & $0.75\pm 0.01$ \\
HR~3498 & 75311 & B3V(n) &$0.6\pm 0.1$ & $0.3\pm 0.1$ & $0.22\pm 0.04$ & $0.6\pm 0.1$ & $0.32\pm 0.03$ & $0.50\pm 0.01$ & $0.203\pm 0.004$ \\
HR~3593 & 77320 & B2Vnn(e) & -- & -- & -- & $0.33\pm 0.04$ & $0.105\pm 0.005$ & $0.317\pm 0.004$ & $0.166\pm 0.004$ \\
HR~3858 & 83953 & B5V &$1.6\pm 0.2$ & $0.9\pm 0.1$ & $0.4\pm 0.2$ &  -- & -- & $1.09\pm 0.01$ & $0.58\pm 0.01$ \\
HD~86612 & 86612 & B5Ve & -- & -- & -- & $0.60\pm 0.04$ & $0.36\pm 0.03$ & $0.44\pm 0.01$ & $0.27\pm 0.01$ \\
HD~88661 & 88661 & B5Vne &$0.8\pm 0.1$ & $0.5\pm 0.1$ & -- & $1.05\pm 0.01$ & $0.58\pm 0.02$ & $0.82\pm 0.01$ & $0.52\pm 0.01$ \\
$\omega$~Car & 89080 & B8IIIe &$2.4\pm 0.2$ & $1.1\pm 0.1$ & $0.50\pm 0.05$ & $3.13\pm 0.02$ & $1.41\pm 0.02$ & $2.27\pm 0.01$ & $1.12\pm 0.01$ \\
HD~91120 & 91120 & B8/9IV/V & -- & -- & -- & $0.45\pm 0.03$ & $0.22\pm 0.02$ & $0.288\pm 0.003$ & $0.143\pm 0.003$ \\
p~Car & 91465 & B4Vne &$5.5\pm 0.3$ & $2.7\pm 0.2$ & -- & $7.0\pm 0.1$ & $3.29\pm 0.02$ & $7.7\pm 0.1$ & $3.65\pm 0.05$ \\
$\delta$~Cen & 105435 & B2Vne &$12.7\pm 0.9$ & $7.1\pm 0.8$ & $3.0\pm 0.4$ & $13.3\pm 0.2$ & $7.0\pm 0.1$ & $11.8\pm 0.1$ & $6.1\pm 0.1$ \\
HD~105521 & 105521 & B3IVe & -- & -- & -- &  -- & -- & $0.46\pm 0.01$ & $0.217\pm 0.004$ \\
$\kappa$~Dra & 109387 & B6IIIe &$3.4\pm 0.2$ & $1.7\pm 0.1$ & $0.5\pm 0.1$ & $4.06\pm 0.03$ & $2.18\pm 0.04$ & $2.83\pm 0.01$ & $1.56\pm 0.02$ \\
HD~110432 & 110432 & B0.5IVpe &$3.8\pm 0.3$ & $2.0\pm 0.3$ & -- & $5.25\pm 0.05$ & $2.42\pm 0.03$ & $3.88\pm 0.03$ & $2.03\pm 0.03$ \\
$\lambda$~Cru & 112078 & B3Vne & -- & -- & -- &  -- & -- & $0.251\pm 0.003$ & $0.076\pm 0.002$ \\
$\mu^2$~Cru & 112091 & B5Vne &$0.9\pm 0.1$ & $0.6\pm 0.1$ & -- &  -- & -- &  -- & -- \\ 
$\mu$~Cen & 120324 & B2Vnpe & -- & -- & -- & $4.7\pm 0.3$ & $1.9\pm 0.1$ & $0.89\pm 0.01$ & $0.46\pm 0.01$ \\
HD~124367 & 124367 & B4Vne &$1.7\pm 0.3$ & $0.8\pm 0.1$ & -- & $1.69\pm 0.02$ & $0.96\pm 0.03$ & $1.25\pm 0.02$ & $0.81\pm 0.01$ \\
$\eta$~Cen & 127972 & B2Ve & -- & -- & -- & $9.2\pm 0.6$ & $4.4\pm 0.1$ & $4.47\pm 0.03$ & $2.39\pm 0.03$ \\
$\theta$~Cir & 131492 & B2IV/V & -- & -- & -- & $0.8\pm 0.1$ & $0.34\pm 0.04$ & $0.77\pm 0.01$ & $0.32\pm 0.01$ \\
$\mu$~Lup & 135734 & B8Ve &$0.9\pm 0.1$ & $0.4\pm 0.1$ & $0.3\pm 0.1$ & $1.25\pm 0.03$ & $0.62\pm 0.01$ & $0.81\pm 0.01$ & $0.41\pm 0.01$ \\
$\chi$~Oph & 148184 & B2Vne &$8.8\pm 0.8$ & $4.5\pm 0.6$ & $2.3\pm 0.4$ & $12.911\pm 0.002$ & $6.45\pm 0.05$ & $10.9\pm 0.2$ & $5.5\pm 0.1$ \\
$\zeta$~Oph & 149757 & O9.2IV & -- & -- & -- & $4.0\pm 0.1$ & $1.06\pm 0.04$ & $2.38\pm 0.03$ & $0.78\pm 0.02$ \\
$\iota$~Ara & 157042 & B2(V)nne &$1.1\pm 0.1$ & $0.6\pm 0.1$ & -- & $1.16\pm 0.03$ & $0.53\pm 0.04$ & $0.83\pm 0.01$ & $0.43\pm 0.01$ \\
$\alpha$~Ara & 158427 & B2Vne &$10.3\pm 0.7$ & $4.9\pm 0.5$ & $1.7\pm 0.2$ &  -- & -- & $10.8\pm 0.2$ & $5.65\pm 0.05$ \\
$66$~Oph & 164284 & B2Ve &$2.9\pm 0.2$ & $1.6\pm 0.2$ & $0.8\pm 0.1$ &  -- & -- &  -- & -- \\ 
$\lambda$~Pav & 173948 & B2Ve & -- & -- & -- &  -- & -- & $0.40\pm 0.01$ & $0.124\pm 0.003$ \\
$64$~Ser & 175869 & B8/9II & -- & -- & -- &  -- & -- & $0.299\pm 0.004$ & $0.180\pm 0.004$ \\
$\beta$~Cyg~B & 183914 & B8Ve & -- & -- & -- &  -- & -- & $0.39\pm 0.01$ & $0.142\pm 0.002$ \\
$11$~Cyg & 185037 & B8Vne & -- & -- & -- &  -- & -- & $0.196\pm 0.003$ & $0.106\pm 0.003$ \\
$12$~Vul & 187811 & B2.5Ve & -- & -- & -- &  -- & -- & $0.366\pm 0.005$ & $0.131\pm 0.003$ \\
$25$~Cyg & 189687 & B3IVe & -- & -- & -- & $0.39\pm 0.01$ & $0.26\pm 0.02$ &  -- & -- \\ 
$28$~Cyg & 191610 & B2.5Ve &$1.0\pm 0.1$ & $0.5\pm 0.1$ & -- & $0.40\pm 0.02$ & $0.17\pm 0.05$ & $0.277\pm 0.004$ & $0.132\pm 0.003$ \\
$20$~Vul & 192044 & B7Ve & -- & -- & -- & $0.36\pm 0.02$ & $0.15\pm 0.01$ & $0.252\pm 0.003$ & $0.141\pm 0.004$ \\
$25$~Vul & 193911 & B6IVe &$0.4\pm 0.1$ & $0.22\pm 0.05$ & -- &  -- & -- &  -- & -- \\ 
HD~194335 & 194335 & B2IIIe & -- & -- & -- & $0.45\pm 0.01$ & $0.3\pm 0.1$ & $0.350\pm 0.005$ & $0.165\pm 0.004$ \\
59~Cyg & 200120 & B1.5Vnne & -- & -- & -- & $2.5\pm 0.1$ & $0.95\pm 0.04$ & $1.46\pm 0.02$ & $0.60\pm 0.01$ \\
$6$~Cep & 203467 & B3IVe & -- & -- & -- & $2.52\pm 0.05$ & $2.4\pm 0.1$ & $1.80\pm 0.02$ & $1.93\pm 0.03$ \\
HD~208682 & 208682 & B2Ve &$0.4\pm 0.1$ & $0.30\pm 0.05$ & -- &  -- & -- &  -- & -- \\ 
$\eta$~PsA & 209014 & B8/9V+B8/9 & -- & -- & -- & $0.507\pm 0.001$ & $0.23\pm 0.03$ & $0.337\pm 0.004$ & $0.182\pm 0.005$ \\
$31$~Peg & 212076 & B2IV-Ve & -- & -- & -- & $2.08\pm 0.03$ & $0.99\pm 0.02$ & $0.81\pm 0.01$ & $0.60\pm 0.01$ \\
$\pi$~Aqr & 212571 & B1III-IVe & -- & -- & -- &  -- & -- & $0.72\pm 0.01$ & $0.31\pm 0.01$ \\
$\epsilon$~PsA & 214748 & B8Ve &$0.9\pm 0.1$ & $0.4\pm 0.1$ & -- & $1.15\pm 0.01$ & $0.52\pm 0.01$ & $0.81\pm 0.01$ & $0.43\pm 0.01$ \\
$\beta$~Psc & 217891 & B6Ve &$0.6\pm 0.1$ & $0.4\pm 0.1$ & $0.3\pm 0.1$ & $1.5\pm 0.2$ & $0.7\pm 0.1$ & $1.48\pm 0.02$ & $0.85\pm 0.01$ \\
HD~224559 & 224559 & B4Vne & -- & -- & -- & $0.42\pm 0.02$ & $0.184\pm 0.003$ & $0.311\pm 0.004$ & $0.197\pm 0.004$ \\
\hline
\noalign{\smallskip}
\multicolumn{10}{l}{(a) Spectral classification provided by SIMBAD Astronomical Database\footnote{\url{http://simbad.u-strasbg.fr/simbad/}}}.
\end{longtable}
\end{landscape}
\twocolumn
}
%-------------------------------------------------

%-------------------------------------------------
% Table fit results
%-------------------------------------------------
%\afterpage{
\nonstopmode
\setlongtables
\onecolumn
\begin{landscape}
\def\arraystretch{1.5}
\begin{longtable}{l|cccc|ccc|ccc|ccc}
\kill
\caption{Stellar and disc parameters sampled with the {\ttfamily emcee} code.} \label{tab:results}
\\
\hline
HD & & & & & & IRAS & & & AKARI & & & WISE &\\
 & $M_{\star}$ & $R_{\mathrm{pole}}$ & $\log L_{\star}$ & $W$ & $n$ & $\log\rho_0$ & $\log(\dot{M}_{\mathrm{SS}}/\alpha)$ & $n$ & $\log\rho_0$ & $\log(\dot{M}_{\mathrm{SS}}/\alpha)$ & $n$ & $\log\rho_0$ & $\log(\dot{M}_{\mathrm{SS}}/\alpha)$ \\
 & $\mathrm{[M_{\odot}]}$ & $\mathrm{[R_{\odot}]}$ & $\mathrm{[L_{\odot}]}$ & & & $[\mathrm{g\,cm^{-3}}]$ & $[\mathrm{M_{\odot}\,yr^{-1}}]$ & $ $ & $[\mathrm{g\,cm^{-3}}]$ & $[\mathrm{M_{\odot}\,yr^{-1}}]$ & $ $ & $[\mathrm{g\,cm^{-3}}]$ & $[\mathrm{M_{\odot}\,yr^{-1}}]$ \\ \hline
5394 & $25\pm 9$ & $8\pm 2$ & $4.7\pm 0.3$ & $0.9\pm 0.1$ & $3.8\substack{+0.9\\-0.6}$ & $-10.8\substack{+0.6\\-0.5}$ & $-9.6$ & $3.29\substack{+0.05\\-0.05}$ & $-10.9\substack{+0.4\\-0.3}$ & $-9.7$ &  -- & -- & -- \\
6811 & $5.6\pm 1.0$ & $10\pm 1$ & $3.3\pm 0.1$ & $0.6\pm 0.2$ & $5\substack{+12\\-2}$ & $-12.4\substack{+1.2\\-0.3}$ & $-10.9$ &  -- & -- & -- &  -- & -- & -- \\
11606 & $9\pm 2$ & $6.2\pm 0.6$ & $3.9\pm 0.1$ & $0.7\pm 0.1$ & $4\substack{+9\\-1}$ & $-11.6\substack{+3.1\\-0.6}$ & $-10.6$ & $2.8\substack{+1.5\\-0.6}$ & $-12.0\substack{+0.5\\-0.2}$ & $-11.0$ & $2.3\substack{+0.1\\-0.2}$ & $-12.2\substack{+0.1\\-0.1}$ & $-11.2$ \\
18552 & $3.9\pm 0.4$ & $3.9\pm 0.2$ & $2.78\pm 0.05$ & $0.80\pm 0.05$ &  -- & -- & -- & $2.1\substack{+0.3\\-0.2}$ & $-12.3\substack{+0.1\\-0.2}$ & $-11.6$ & $2.1\substack{+0.1\\-0.1}$ & $-12.4\substack{+0.1\\-0.1}$ & $-11.7$ \\
20336 & $7\pm 1$ & $3.9\pm 0.4$ & $3.4\pm 0.1$ & $0.7\pm 0.1$ &  -- & -- & -- & $3.3\substack{+0.3\\-0.2}$ & $-10.8\substack{+0.3\\-0.3}$ & $-10.3$ & $2.6\substack{+0.1\\-0.1}$ & $-11.2\substack{+0.1\\-0.1}$ & $-10.8$ \\
22780 & $6.4\pm 0.8$ & $6.5\pm 0.5$ & $3.3\pm 0.1$ & $0.8\pm 0.1$ &  -- & -- & -- &  -- & -- & -- & $2.5\substack{+0.3\\-0.3}$ & $-12.3\substack{+0.2\\-0.2}$ & $-11.2$ \\
23016 & $3.5\pm 0.6$ & $2.7\pm 0.3$ & $2.3\pm 0.1$ & $0.7\pm 0.1$ &  -- & -- & -- &  -- & -- & -- & $9\substack{+5\\-3}$ & $-11.4\substack{+1.0\\-0.7}$ & $-11.4$ \\
23480 & $3.8\pm 0.6$ & $3.5\pm 0.3$ & $2.7\pm 0.1$ & $0.8\pm 0.1$ &  -- & -- & -- & $2.0\substack{+0.2\\-0.1}$ & $-12.1\substack{+0.1\\-0.1}$ & $-11.6$ & $2.0\substack{+0.1\\-0.1}$ & $-12.2\substack{+0.1\\-0.1}$ & $-11.8$ \\
23552 & $4.5\pm 0.7$ & $3.9\pm 0.4$ & $2.7\pm 0.1$ & $0.7\pm 0.1$ &  -- & -- & -- & $2.4\substack{+0.4\\-0.3}$ & $-11.9\substack{+0.2\\-0.2}$ & $-11.4$ & $2.5\substack{+0.1\\-0.1}$ & $-11.9\substack{+0.2\\-0.2}$ & $-11.5$ \\
23630 & $5\pm 1$ & $9\pm 1$ & $3.3\pm 0.1$ & $0.8\pm 0.1$ & $4\substack{+4\\-1}$ & $-12.3\substack{+0.9\\-0.3}$ & $-10.6$ & $2.9\substack{+0.2\\-0.3}$ & $-12.3\substack{+0.2\\-0.1}$ & $-10.6$ & $3.2\substack{+0.2\\-0.3}$ & $-12.3\substack{+0.3\\-0.2}$ & $-10.7$ \\
24534 & $11\pm 2$ & $6.1\pm 0.7$ & $4.2\pm 0.1$ & $0.7\pm 0.1$ &  -- & -- & -- & $4.0\substack{+0.3\\-0.2}$ & $-10.3\substack{+0.4\\-0.3}$ & $-9.2$ & $3.1\substack{+0.1\\-0.1}$ & $-10.9\substack{+0.3\\-0.3}$ & $-9.8$ \\
25940 & $5.5\pm 0.8$ & $4.8\pm 0.5$ & $3.3\pm 0.1$ & $0.7\pm 0.2$ & $2.9\substack{+0.2\\-0.2}$ & $-11.4\substack{+0.2\\-0.2}$ & $-10.6$ & $2.8\substack{+0.1\\-0.1}$ & $-11.4\substack{+0.1\\-0.1}$ & $-10.7$ & $2.7\substack{+0.1\\-0.1}$ & $-11.5\substack{+0.1\\-0.1}$ & $-10.7$ \\
28497 & $14\pm 4$ & $4.7\pm 0.9$ & $4.1\pm 0.2$ & $0.7\pm 0.1$ & $4\substack{+6\\-1}$ & $-10.3\substack{+3.3\\-0.8}$ & $-9.7$ & $4.0\substack{+2.1\\-0.8}$ & $-10.5\substack{+0.9\\-0.5}$ & $-9.9$ & $3.4\substack{+0.1\\-0.1}$ & $-10.7\substack{+0.3\\-0.3}$ & $-10.2$ \\
32343 & $6.1\pm 1.0$ & $4.8\pm 0.6$ & $3.1\pm 0.1$ & $0.6\pm 0.2$ & $1.7\substack{+0.1\\-0.1}$ & $-12.2\substack{+0.1\\-0.1}$ & $-11.7$ & $2.2\substack{+0.2\\-0.2}$ & $-12.0\substack{+0.1\\-0.1}$ & $-11.4$ & $2.0\substack{+0.1\\-0.1}$ & $-12.1\substack{+0.1\\-0.1}$ & $-11.5$ \\
33328 & $10\pm 1$ & $7.4\pm 0.6$ & $4.2\pm 0.1$ & $0.8\pm 0.1$ &  -- & -- & -- &  -- & -- & -- & $2.9\substack{+0.4\\-0.4}$ & $-12.2\substack{+0.3\\-0.2}$ & $-10.9$ \\
35439 & $11\pm 2$ & $5.5\pm 0.5$ & $3.9\pm 0.1$ & $0.7\pm 0.1$ & $6\substack{+11\\-2}$ & $-10\substack{+4\\-1}$ & $-8.7$ & $4.2\substack{+1.4\\-0.7}$ & $-10.6\substack{+0.6\\-0.4}$ & $-9.8$ & $3.4\substack{+0.1\\-0.1}$ & $-11.0\substack{+0.3\\-0.3}$ & $-10.2$ \\
36576 & $11\pm 2$ & $6.2\pm 0.6$ & $4.1\pm 0.1$ & $0.7\pm 0.1$ & $1.7\substack{+0.1\\-0.1}$ & $-12.0\substack{+0.2\\-0.2}$ & $-11.0$ & $2.3\substack{+0.1\\-0.1}$ & $-11.4\substack{+0.1\\-0.1}$ & $-10.4$ & $2.36\substack{+0.04\\-0.04}$ & $-11.4\substack{+0.1\\-0.1}$ & $-10.4$ \\
37657 & $10\pm 1$ & $7.1\pm 0.6$ & $3.8\pm 0.1$ & $0.7\pm 0.1$ &  -- & -- & -- & $2.8\substack{+0.3\\-0.3}$ & $-11.8\substack{+0.3\\-0.3}$ & $-10.7$ & $2.0\substack{+0.1\\-0.1}$ & $-12.2\substack{+0.1\\-0.1}$ & $-11.1$ \\
37795 & $5.7\pm 0.6$ & $6.0\pm 0.4$ & $3.0\pm 0.1$ & $0.9\pm 0.1$ & $2.9\substack{+0.2\\-0.2}$ & $-12.2\substack{+0.1\\-0.1}$ & $-11.1$ & $2.5\substack{+0.1\\-0.2}$ & $-12.2\substack{+0.1\\-0.1}$ & $-11.2$ & $2.5\substack{+0.2\\-0.2}$ & $-12.3\substack{+0.1\\-0.1}$ & $-11.3$ \\
37967 & $8\pm 2$ & $5.0\pm 0.8$ & $3.3\pm 0.1$ & $0.7\pm 0.1$ &  -- & -- & -- & $2.6\substack{+0.1\\-0.1}$ & $-11.8\substack{+0.2\\-0.2}$ & $-11.1$ & $2.4\substack{+0.1\\-0.1}$ & $-11.8\substack{+0.2\\-0.2}$ & $-11.2$ \\
38010 & $13\pm 3$ & $4.4\pm 0.7$ & $4.0\pm 0.1$ & $0.7\pm 0.1$ & $2.3\substack{+1.3\\-0.4}$ & $-11.5\substack{+0.9\\-0.4}$ & $-10.9$ & $2.4\substack{+0.5\\-0.3}$ & $-11.5\substack{+0.3\\-0.4}$ & $-10.9$ & $2.5\substack{+0.1\\-0.1}$ & $-11.5\substack{+0.2\\-0.2}$ & $-11.0$ \\
40978 & $9\pm 2$ & $7\pm 1$ & $3.8\pm 0.1$ & $0.7\pm 0.1$ &  -- & -- & -- &  -- & -- & -- & $2.7\substack{+0.2\\-0.2}$ & $-12.0\substack{+0.3\\-0.2}$ & $-10.7$ \\
41335 & $8\pm 2$ & $4.1\pm 0.7$ & $3.6\pm 0.1$ & $0.7\pm 0.1$ & $3.4\substack{+1.4\\-0.6}$ & $-9.8\substack{+1.4\\-0.7}$ & $-9.2$ &  -- & -- & -- & $3.0\substack{+0.1\\-0.1}$ & $-10.3\substack{+0.3\\-0.3}$ & $-9.8$ \\
44458 & $13\pm 3$ & $8.4\pm 1.0$ & $4.4\pm 0.1$ & $0.7\pm 0.1$ & $4\substack{+5\\-1}$ & $-11.0\substack{+2.0\\-0.7}$ & $-9.6$ & $3.2\substack{+0.9\\-0.4}$ & $-11.7\substack{+0.4\\-0.4}$ & $-10.3$ & $2.4\substack{+0.1\\-0.1}$ & $-12.0\substack{+0.2\\-0.2}$ & $-10.7$ \\
45995 & $10\pm 2$ & $5.9\pm 0.6$ & $3.9\pm 0.1$ & $0.7\pm 0.1$ &  -- & -- & -- & $6\substack{+5\\-2}$ & $-11.2\substack{+1.2\\-0.8}$ & $-10.2$ & $2.4\substack{+0.1\\-0.1}$ & $-12.0\substack{+0.1\\-0.1}$ & $-11.1$ \\
47054 & $5.0\pm 0.8$ & $5.3\pm 0.6$ & $2.9\pm 0.1$ & $0.7\pm 0.1$ &  -- & -- & -- & $5\substack{+14\\-2}$ & $-11.8\substack{+2.1\\-0.5}$ & $-11.1$ & $2.4\substack{+0.2\\-0.2}$ & $-12.3\substack{+0.2\\-0.2}$ & $-11.5$ \\
50013 & $12\pm 3$ & $5.3\pm 1.0$ & $4.0\pm 0.2$ & $0.7\pm 0.2$ & $3.5\substack{+0.3\\-0.3}$ & $-10.7\substack{+0.4\\-0.3}$ & $-9.9$ & $3.2\substack{+0.3\\-0.2}$ & $-11.0\substack{+0.3\\-0.3}$ & $-10.2$ & $3.4\substack{+0.1\\-0.1}$ & $-10.8\substack{+0.3\\-0.3}$ & $-10.0$ \\
56139 & $10\pm 4$ & $8\pm 2$ & $3.9\pm 0.2$ & $0.6\pm 0.2$ & $3.2\substack{+0.8\\-0.5}$ & $-11.6\substack{+0.5\\-0.4}$ & $-10.4$ & $3.2\substack{+3.1\\-0.9}$ & $-11.4\substack{+1.3\\-0.5}$ & $-10.2$ & $3.3\substack{+0.1\\-0.1}$ & $-11.5\substack{+0.3\\-0.3}$ & $-10.3$ \\
58050 & $10\pm 2$ & $5.6\pm 0.9$ & $3.6\pm 0.1$ & $0.6\pm 0.2$ &  -- & -- & -- &  -- & -- & -- & $6\substack{+3\\-1}$ & $-11\substack{+1\\-1}$ & $-10.1$ \\
58343 & $7\pm 1$ & $6.5\pm 0.9$ & $3.4\pm 0.1$ & $0.5\pm 0.2$ & $2.8\substack{+2.8\\-0.6}$ & $-11.7\substack{+1.2\\-0.4}$ & $-10.8$ &  -- & -- & -- & $2.1\substack{+0.1\\-0.1}$ & $-12.2\substack{+0.1\\-0.1}$ & $-11.4$ \\
58715 & $3.9\pm 0.5$ & $3.6\pm 0.3$ & $2.5\pm 0.1$ & $0.8\pm 0.1$ & $2.9\substack{+0.3\\-0.3}$ & $-12.1\substack{+0.2\\-0.2}$ & $-11.7$ & $3.0\substack{+1.1\\-0.3}$ & $-12.1\substack{+0.1\\-0.1}$ & $-11.7$ & $2.7\substack{+0.2\\-0.2}$ & $-12.1\substack{+0.1\\-0.1}$ & $-11.8$ \\
58978 & $13\pm 3$ & $4.5\pm 0.8$ & $3.9\pm 0.1$ & $0.7\pm 0.1$ & $11\substack{+47\\-6}$ & $-7\substack{+23\\-3}$ & $-6.2$ &  -- & -- & -- & $3.1\substack{+0.2\\-0.2}$ & $-11.6\substack{+0.3\\-0.3}$ & $-11.1$ \\
60606 & $9\pm 2$ & $6.3\pm 0.7$ & $3.7\pm 0.1$ & $0.7\pm 0.1$ & $2.1\substack{+0.1\\-0.1}$ & $-11.8\substack{+0.2\\-0.2}$ & $-10.8$ & $2.2\substack{+0.1\\-0.1}$ & $-11.8\substack{+0.1\\-0.1}$ & $-10.8$ & $2.5\substack{+0.1\\-0.1}$ & $-11.7\substack{+0.1\\-0.1}$ & $-10.8$ \\
60848 & $14\pm 3$ & $5.5\pm 0.6$ & $4.3\pm 0.1$ & $0.7\pm 0.2$ &  -- & -- & -- &  -- & -- & -- & $3.0\substack{+0.2\\-0.2}$ & $-11.9\substack{+0.2\\-0.2}$ & $-11.1$ \\
63462 & $20\pm 3$ & $8.0\pm 0.8$ & $4.7\pm 0.1$ & $0.94\pm 0.06$ & $6\substack{+6\\-2}$ & $-9\substack{+4\\-1}$ & $-7.5$ &  -- & -- & -- & $4.6\substack{+0.2\\-0.2}$ & $-10.1\substack{+0.3\\-0.3}$ & $-8.7$ \\
65875 & $8\pm 2$ & $5.2\pm 0.6$ & $3.6\pm 0.1$ & $0.7\pm 0.2$ &  -- & -- & -- & $2.7\substack{+0.2\\-0.2}$ & $-11.2\substack{+0.2\\-0.2}$ & $-10.4$ & $2.5\substack{+0.1\\-0.1}$ & $-11.3\substack{+0.2\\-0.2}$ & $-10.5$ \\
68980 & $13\pm 2$ & $5.9\pm 0.6$ & $4.1\pm 0.1$ & $0.6\pm 0.2$ & $2.1\substack{+0.1\\-0.1}$ & $-11.8\substack{+0.1\\-0.1}$ & $-11.0$ & $2.8\substack{+0.1\\-0.1}$ & $-11.5\substack{+0.1\\-0.2}$ & $-10.7$ & $2.8\substack{+0.1\\-0.1}$ & $-11.5\substack{+0.1\\-0.1}$ & $-10.7$ \\
75311 & $5\pm 1$ & $3.3\pm 0.6$ & $3.0\pm 0.1$ & $0.7\pm 0.1$ & $2.3\substack{+0.2\\-0.2}$ & $-11.7\substack{+0.2\\-0.2}$ & $-11.5$ & $3.0\substack{+5.0\\-0.8}$ & $-11.5\substack{+1.7\\-0.5}$ & $-11.2$ & $4.3\substack{+0.3\\-0.3}$ & $-11.0\substack{+0.3\\-0.4}$ & $-10.8$ \\
77320 & $9\pm 2$ & $4.7\pm 0.9$ & $3.6\pm 0.2$ & $0.7\pm 0.1$ &  -- & -- & -- & $10\substack{+14\\-5}$ & $-9\substack{+4\\-2}$ & $-8.7$ & $2.8\substack{+0.1\\-0.2}$ & $-11.6\substack{+0.3\\-0.3}$ & $-11.1$ \\
83953 & $6.8\pm 1.0$ & $5.3\pm 0.5$ & $3.3\pm 0.1$ & $0.7\pm 0.1$ & $2.8\substack{+0.7\\-0.4}$ & $-11.7\substack{+0.4\\-0.2}$ & $-10.9$ &  -- & -- & -- & $2.6\substack{+0.1\\-0.1}$ & $-12.0\substack{+0.1\\-0.1}$ & $-11.2$ \\
86612 & $5.3\pm 0.8$ & $4.7\pm 0.5$ & $3.3\pm 0.1$ & $0.7\pm 0.1$ &  -- & -- & -- & $2.4\substack{+0.5\\-0.3}$ & $-12.0\substack{+0.2\\-0.2}$ & $-11.2$ & $2.4\substack{+0.1\\-0.1}$ & $-12.0\substack{+0.1\\-0.2}$ & $-11.2$ \\
88661 & $10\pm 2$ & $4.9\pm 0.8$ & $3.7\pm 0.1$ & $0.7\pm 0.2$ & $4\substack{+13\\-2}$ & $-10\substack{+6\\-1}$ & $-9.6$ & $2.7\substack{+0.1\\-0.1}$ & $-11.3\substack{+0.2\\-0.2}$ & $-10.6$ & $2.5\substack{+0.1\\-0.1}$ & $-11.4\substack{+0.2\\-0.2}$ & $-10.8$ \\
89080 & $4.0\pm 0.6$ & $5.1\pm 0.5$ & $2.8\pm 0.1$ & $0.8\pm 0.1$ & $2.9\substack{+0.2\\-0.2}$ & $-11.8\substack{+0.2\\-0.2}$ & $-10.9$ & $3.0\substack{+0.2\\-0.2}$ & $-11.7\substack{+0.2\\-0.2}$ & $-10.8$ & $3.0\substack{+0.1\\-0.1}$ & $-11.8\substack{+0.2\\-0.2}$ & $-10.9$ \\
91120 & $2.6\pm 0.4$ & $2.1\pm 0.2$ & $2.0\pm 0.1$ & $0.74\pm 0.06$ &  -- & -- & -- & $3.5\substack{+1.3\\-0.6}$ & $-10.7\substack{+0.9\\-0.4}$ & $-11.0$ & $3.1\substack{+0.1\\-0.1}$ & $-11.0\substack{+0.2\\-0.2}$ & $-11.3$ \\
91465 & $7\pm 2$ & $5.8\pm 1.0$ & $3.6\pm 0.1$ & $0.7\pm 0.1$ & $3.0\substack{+0.5\\-0.3}$ & $-11.1\substack{+0.4\\-0.3}$ & $-10.1$ & $3.1\substack{+0.1\\-0.1}$ & $-11.0\substack{+0.2\\-0.3}$ & $-10.0$ & $3.3\substack{+0.1\\-0.1}$ & $-10.7\substack{+0.3\\-0.2}$ & $-9.7$ \\
105435 & $12\pm 2$ & $5.7\pm 0.6$ & $3.9\pm 0.1$ & $0.7\pm 0.2$ & $2.8\substack{+0.2\\-0.2}$ & $-11.2\substack{+0.2\\-0.2}$ & $-10.4$ & $2.8\substack{+0.1\\-0.1}$ & $-11.3\substack{+0.1\\-0.1}$ & $-10.5$ & $3.0\substack{+0.1\\-0.1}$ & $-11.1\substack{+0.2\\-0.1}$ & $-10.3$ \\
105521 & $10\pm 1$ & $12.7\pm 1.0$ & $4.1\pm 0.1$ & $0.7\pm 0.2$ &  -- & -- & -- &  -- & -- & -- & $2.9\substack{+0.2\\-0.4}$ & $-12.4\substack{+0.1\\-0.1}$ & $-10.5$ \\
109387 & $7\pm 1$ & $6.9\pm 0.7$ & $3.4\pm 0.1$ & $0.52\pm 0.03$ & $3.2\substack{+0.3\\-0.2}$ & $-11.2\substack{+0.3\\-0.2}$ & $-10.3$ & $2.5\substack{+0.1\\-0.1}$ & $-11.7\substack{+0.2\\-0.2}$ & $-10.8$ & $2.6\substack{+0.1\\-0.1}$ & $-11.7\substack{+0.1\\-0.2}$ & $-10.8$ \\
110432 & $9\pm 2$ & $4.8\pm 0.6$ & $3.8\pm 0.1$ & $0.9\pm 0.1$ & $3.1\substack{+0.9\\-0.5}$ & $-10.0\substack{+1.0\\-0.6}$ & $-9.1$ & $3.3\substack{+0.1\\-0.1}$ & $-9.9\substack{+0.2\\-0.2}$ & $-9.0$ & $3.0\substack{+0.1\\-0.1}$ & $-10.2\substack{+0.2\\-0.2}$ & $-9.3$ \\
112078 & $4.7\pm 0.5$ & $2.5\pm 0.2$ & $2.78\pm 0.07$ & $0.7\pm 0.1$ &  -- & -- & -- &  -- & -- & -- & $11\substack{+7\\-5}$ & $-11.7\substack{+1.2\\-0.5}$ & $-11.7$ \\
112091 & $7.9\pm 1.3$ & $4.5\pm 0.4$ & $3.6\pm 0.1$ & $0.7\pm 0.2$ & $2.6\substack{+1.0\\-0.5}$ & $-12.2\substack{+0.2\\-0.2}$ & $-11.5$ &  -- & -- & -- &  -- & -- & -- \\
120324 & $10\pm 2$ & $5.3\pm 0.6$ & $3.8\pm 0.1$ & $0.7\pm 0.1$ &  -- & -- & -- & $4.2\substack{+1.2\\-0.7}$ & $-11.3\substack{+0.5\\-0.3}$ & $-10.5$ & $2.6\substack{+0.2\\-0.2}$ & $-12.1\substack{+0.1\\-0.1}$ & $-11.5$ \\
124367 & $6.3\pm 1.0$ & $4.1\pm 0.4$ & $3.3\pm 0.1$ & $0.7\pm 0.1$ & $5\substack{+11\\-2}$ & $-10\substack{+5\\-1}$ & $-9.7$ & $2.4\substack{+0.1\\-0.1}$ & $-11.8\substack{+0.1\\-0.1}$ & $-11.2$ & $2.3\substack{+0.1\\-0.1}$ & $-11.9\substack{+0.1\\-0.1}$ & $-11.3$ \\
127972 & $10\pm 2$ & $5.2\pm 0.6$ & $3.7\pm 0.1$ & $0.7\pm 0.1$ &  -- & -- & -- & $3.1\substack{+0.4\\-0.3}$ & $-11.4\substack{+0.3\\-0.2}$ & $-10.7$ & $2.7\substack{+0.1\\-0.1}$ & $-11.8\substack{+0.1\\-0.2}$ & $-11.1$ \\
131492 & $10\pm 2$ & $8.0\pm 0.8$ & $3.9\pm 0.1$ & $0.7\pm 0.2$ &  -- & -- & -- & $8\substack{+22\\-4}$ & $-10\substack{+8\\-1}$ & $-8.7$ & $4.2\substack{+0.2\\-0.2}$ & $-11.1\substack{+0.2\\-0.2}$ & $-9.8$ \\
135734 & $3.2\pm 0.8$ & $2.7\pm 0.5$ & $2.5\pm 0.2$ & $0.7\pm 0.1$ & $2.4\substack{+0.3\\-0.2}$ & $-11.7\substack{+0.3\\-0.3}$ & $-11.6$ & $2.8\substack{+0.2\\-0.2}$ & $-11.5\substack{+0.3\\-0.3}$ & $-11.4$ & $2.9\substack{+0.2\\-0.2}$ & $-11.5\substack{+0.3\\-0.3}$ & $-11.4$ \\
148184 & $21\pm 4$ & $7.9\pm 1.2$ & $4.6\pm 0.1$ & $0.6\pm 0.2$ & $2.7\substack{+0.3\\-0.2}$ & $-11.6\substack{+0.2\\-0.2}$ & $-10.5$ & $2.9\substack{+1.1\\-0.1}$ & $-11.5\substack{+0.2\\-0.1}$ & $-10.4$ & $3.1\substack{+0.1\\-0.1}$ & $-11.3\substack{+0.2\\-0.2}$ & $-10.3$ \\
149757 & $11\pm 2$ & $4.9\pm 0.5$ & $4.2\pm 0.1$ & $0.7\pm 0.1$ &  -- & -- & -- & $18\substack{+26\\-7}$ & $-8\substack{+5\\-2}$ & $-7.7$ & $7\substack{+2\\-1}$ & $-10.9\substack{+0.7\\-0.6}$ & $-10.1$ \\
157042 & $8\pm 2$ & $3.0\pm 0.6$ & $3.4\pm 0.2$ & $0.7\pm 0.1$ & $3.0\substack{+1.2\\-0.5}$ & $-10.6\substack{+1.0\\-0.5}$ & $-10.4$ & $3.5\substack{+0.5\\-0.4}$ & $-10.4\substack{+0.4\\-0.4}$ & $-10.3$ & $2.9\substack{+0.1\\-0.1}$ & $-10.8\substack{+0.2\\-0.2}$ & $-10.6$ \\
158427 & $4.1\pm 0.6$ & $2.3\pm 0.2$ & $2.8\pm 0.1$ & $0.9\pm 0.1$ & $3.3\substack{+0.3\\-0.3}$ & $-10.3\substack{+0.3\\-0.3}$ & $-10.2$ &  -- & -- & -- & $2.9\substack{+0.1\\-0.1}$ & $-10.5\substack{+0.1\\-0.1}$ & $-10.4$ \\
164284 & $10\pm 3$ & $4.8\pm 0.9$ & $3.7\pm 0.2$ & $0.7\pm 0.1$ & $2.6\substack{+0.1\\-0.1}$ & $-11.3\substack{+0.2\\-0.2}$ & $-10.6$ &  -- & -- & -- &  -- & -- & -- \\
173948 & $12\pm 2$ & $9\pm 1$ & $4.2\pm 0.1$ & $0.7\pm 0.2$ &  -- & -- & -- &  -- & -- & -- & $10\substack{+3\\-3}$ & $-11\substack{+1\\-1}$ & $-10.0$ \\
175869 & $4.6\pm 0.9$ & $6.2\pm 0.7$ & $3.0\pm 0.1$ & $0.7\pm 0.1$ &  -- & -- & -- &  -- & -- & -- & $2.4\substack{+0.1\\-0.1}$ & $-12.1\substack{+0.1\\-0.2}$ & $-11.1$ \\
183914 & $4.0\pm 0.4$ & $3.6\pm 0.3$ & $2.7\pm 0.1$ & $0.7\pm 0.1$ &  -- & -- & -- &  -- & -- & -- & $4.5\substack{+0.7\\-0.9}$ & $-12.1\substack{+0.3\\-0.2}$ & $-11.6$ \\
185037 & $3.9\pm 0.7$ & $3.0\pm 0.4$ & $2.4\pm 0.1$ & $0.7\pm 0.1$ &  -- & -- & -- &  -- & -- & -- & $2.5\substack{+0.2\\-0.2}$ & $-12.0\substack{+0.2\\-0.3}$ & $-11.9$ \\
187811 & $6.5\pm 0.7$ & $4.6\pm 0.3$ & $3.44\pm 0.06$ & $0.7\pm 0.1$ &  -- & -- & -- &  -- & -- & -- & $5\substack{+1\\-1}$ & $-12.1\substack{+0.2\\-0.2}$ & $-11.4$ \\
189687 & $7\pm 2$ & $7.1\pm 0.9$ & $3.8\pm 0.1$ & $0.8\pm 0.1$ &  -- & -- & -- & $2.2\substack{+0.2\\-0.2}$ & $-12.1\substack{+0.1\\-0.1}$ & $-10.8$ &  -- & -- & -- \\
191610 & $5\pm 1$ & $3.7\pm 0.6$ & $3.3\pm 0.2$ & $0.8\pm 0.1$ & $4\substack{+4\\-1}$ & $-10\substack{+3\\-1}$ & $-9.1$ & $10\substack{+26\\-6}$ & $-9\substack{+9\\-2}$ & $-8.0$ & $3.2\substack{+0.2\\-0.2}$ & $-11.2\substack{+0.3\\-0.3}$ & $-10.7$ \\
192044 & $4.2\pm 0.5$ & $5.2\pm 0.3$ & $3.0\pm 0.1$ & $0.76\pm 0.05$ &  -- & -- & -- & $3.1\substack{+2.3\\-0.7}$ & $-12.1\substack{+0.6\\-0.3}$ & $-11.2$ & $2.2\substack{+0.2\\-0.1}$ & $-12.4\substack{+0.1\\-0.1}$ & $-11.4$ \\
193911 & $7\pm 1$ & $11\pm 1$ & $3.5\pm 0.1$ & $0.7\pm 0.1$ & $13\substack{+53\\-9}$ & $-12\substack{+7\\-1}$ & $-10.4$ &  -- & -- & -- &  -- & -- & -- \\
194335 & $11\pm 2$ & $4.4\pm 0.5$ & $3.8\pm 0.1$ & $0.7\pm 0.1$ &  -- & -- & -- & $4\substack{+6\\-2}$ & $-10.7\substack{+2.3\\-0.7}$ & $-10.3$ & $3.3\substack{+0.2\\-0.1}$ & $-11.0\substack{+0.2\\-0.2}$ & $-10.6$ \\
200120 & $12\pm 2$ & $5.8\pm 0.7$ & $4.1\pm 0.1$ & $0.7\pm 0.1$ &  -- & -- & -- & $4.9\substack{+0.8\\-0.6}$ & $-9.0\substack{+0.7\\-0.5}$ & $-8.2$ & $4.4\substack{+0.2\\-0.2}$ & $-9.5\substack{+0.4\\-0.4}$ & $-8.7$ \\
203467 & $8\pm 2$ & $9\pm 1$ & $3.8\pm 0.1$ & $0.7\pm 0.2$ &  -- & -- & -- & $1.8\substack{+0.1\\-0.1}$ & $-12.0\substack{+0.1\\-0.1}$ & $-10.4$ & $1.70\substack{+0.04\\-0.03}$ & $-12.09\substack{+0.05\\-0.04}$ & $-10.7$ \\
208682 & $11\pm 3$ & $5.1\pm 1.0$ & $3.9\pm 0.2$ & $0.7\pm 0.1$ & $2.8\substack{+6.5\\-0.8}$ & $-11.5\substack{+2.4\\-0.5}$ & $-10.8$ &  -- & -- & -- &  -- & -- & -- \\
209014 & $7\pm 2$ & $6\pm 1$ & $3.2\pm 0.2$ & $0.9\pm 0.1$ &  -- & -- & -- & $6\substack{+2\\-2}$ & $-11.2\substack{+0.9\\-0.8}$ & $-10.1$ & $2.5\substack{+0.2\\-0.3}$ & $-12.0\substack{+0.3\\-0.3}$ & $-11.0$ \\
212076 & $10\pm 2$ & $6.4\pm 0.7$ & $3.7\pm 0.1$ & $0.7\pm 0.2$ &  -- & -- & -- & $3.1\substack{+0.1\\-0.1}$ & $-10.8\substack{+0.2\\-0.2}$ & $-9.9$ & $2.13\substack{+0.04\\-0.03}$ & $-11.7\substack{+0.1\\-0.1}$ & $-10.8$ \\
212571 & $14\pm 2$ & $6.2\pm 0.5$ & $4.3\pm 0.1$ & $0.7\pm 0.2$ &  -- & -- & -- &  -- & -- & -- & $3.5\substack{+0.4\\-0.7}$ & $-12.1\substack{+0.2\\-0.2}$ & $-11.1$ \\
214748 & $3.4\pm 0.8$ & $4.4\pm 0.8$ & $2.7\pm 0.1$ & $0.7\pm 0.1$ & $8\substack{+36\\-5}$ & $-10\substack{+13\\-2}$ & $-9.0$ & $3.1\substack{+0.3\\-0.3}$ & $-11.7\substack{+0.3\\-0.3}$ & $-11.0$ & $2.7\substack{+0.1\\-0.2}$ & $-11.9\substack{+0.2\\-0.3}$ & $-11.1$ \\
217891 & $5.3\pm 0.9$ & $5.5\pm 0.5$ & $3.1\pm 0.1$ & $0.32\pm 0.06$ & $2.0\substack{+0.2\\-0.2}$ & $-12.5\substack{+0.2\\-0.2}$ & $-12.0$ & $7\substack{+20\\-4}$ & $-11\substack{+4\\-1}$ & $-10.6$ & $2.4\substack{+0.1\\-0.1}$ & $-12.1\substack{+0.1\\-0.2}$ & $-11.6$ \\
224559 & $7\pm 1$ & $4.8\pm 0.6$ & $3.3\pm 0.1$ & $0.69\pm 0.05$ &  -- & -- & -- & $3.6\substack{+0.6\\-0.4}$ & $-10.6\substack{+0.5\\-0.3}$ & $-9.9$ & $2.4\substack{+0.1\\-0.1}$ & $-11.5\substack{+0.2\\-0.2}$ & $-10.8$ \\
\hline
\noalign{\smallskip}
\end{longtable}
\end{landscape}
\def\arraystretch{1}
\twocolumn
%}
%-------------------------------------------------

%-------------------------------------------------
% build-up and dissipation
%-------------------------------------------------
%\afterpage{
%}
%-------------------------------------------------

\bsp

\label{lastpage}

\end{document}